\documentclass[%
reprint,
superscriptaddress,
 amsmath,amssymb,
 aps,
]{revtex4-2}

\usepackage{newtxtext}
\usepackage{newtxmath}
\usepackage{natbib}
\usepackage{hyperref}
\usepackage{upgreek}
\usepackage{xcolor}
\usepackage{graphicx}% Include figure files
\usepackage{dcolumn}% Align table columns on decimal point
\usepackage{bm}% bold math
\usepackage{orcidlink}
\begin{document}

\preprint{APS/123-QED}

\title{Stagnation points at grain contacts generate an elastic flow instability in 3D porous media}

% \author{Emily Y. Chen\aff{1}, Christopher A. Browne\aff{1}, Simon J. Haward\aff{2}, Amy Q. Shen\aff{2}, \\ \and Sujit S. Datta\aff{1,3}
%   \corresp{\email{ssdatta@caltech.edu}}}

% \affiliation{\aff{1} Department of Chemical and Biological Engineering, Princeton University, Princeton, NJ 08544, USA
% \aff{2} Micro/Bio/Nanofluidics Unit, Okinawa Institute of Science and Technology Graduate University, Onna-son, Okinawa 904-0495, Japan
% \aff{3} Division of Chemistry and Chemical Engineering, California Institute of Technology, Pasadena, CA 91125, USA}

\author{Emily Y. Chen,\orcidlink{0009-0004-6007-3460}}
\author{Christopher A. Browne,\orcidlink{0000-0002-3945-9906}}
\affiliation{%
 Department of Chemical and Biological Engineering, Princeton University, Princeton, NJ 08544, USA
}%

\author{Simon J. Haward,\orcidlink{0000-0002-1884-4100}}
\author{Amy Q. Shen,\orcidlink{0000-0002-1222-6264}}
\affiliation{
 Micro/Bio/Nanofluidics Unit, Okinawa Institute of Science and Technology Graduate University, Onna-son, Okinawa 904-0495, Japan
}%

\author{Sujit S. Datta,\orcidlink{0000-0003-2400-1561}}%
 \email{ssdatta@caltech.edu}
\affiliation{%
 Department of Chemical and Biological Engineering, Princeton University, Princeton, NJ 08544, USA
 }%
\affiliation{%
 Division of Chemistry and Chemical Engineering, California Institute of Technology, Pasadena, CA 91125, USA
 }%

\date{\today}% It is always \today, today,

\begin{abstract}
Many environmental, energy, and industrial processes involve the flow of viscoelastic polymer solutions in three-dimensional (3D) porous media where fluid is confined to navigate through complex pore space geometries. As polymers are transported through the tortuous pore space, elastic stresses accumulate, leading to the onset of unsteady, time-dependent flow fluctuations above a threshold flow rate. How does pore space geometry influence the development and features of this elastic instability? Here, we address this question by directly imaging polymer solution flow in microfabricated 3D ordered porous media with precisely controlled geometries consisting of simple-cubic (SC) or body-centered cuboid (BC) arrays of spherical grains. In both cases, we find that the flow instability is generated at stagnation points arising at the contacts between grains rather than at the polar upstream/downstream grain surfaces, as is the case for flow around a single grain. The characteristics of the flow instability are strongly dependent on the unit cell geometry:  in SC packings, the instability manifests through the formation of time-dependent, fluctuating 3D eddies, whereas in BC packings, it manifests as continual fluctuating ‘wobbles’ and crossing in the flow pathlines. Despite this difference, we find that characteristics of the transition from steady to unsteady flow with increasing flow rate have commonalities across geometries. Moreover, for both packing geometries, our data indicate that extensional flow-induced polymeric stresses generated by contact-associated stagnation points are the primary contributor to the macroscopic resistance to flow across the entire medium. Altogether, our work highlights the pivotal role of inter-grain contacts---which are typically idealized as discrete points and therefore overlooked, but are inherent in most natural and engineered media---in shaping elastic instabilities in porous media.
\end{abstract}

\maketitle

%\tableofcontents

\section{Introduction} \label{sec:Introduction}
The flow of polymer solutions through porous media is widely used across many applications, ranging from groundwater remediation and enhanced oil recovery to filtration and polymer processing~\citep{boudreaux_polymer_1976,bourgeat_filtration_2003,smith_compatibility_2008,sorbie_polymer-improved_2013,pogaku_polymer_2018,firozjaii_review_2020}. In these applications, a polymeric fluid must navigate tortuous flow paths set by the geometry of the pore space. If the flow is sufficiently fast, the constituent polymer molecules become stretched as they are transported along the curved streamlines, and the elastic stresses thereby generated develop and give rise to a flow instability~\citep{larson_instabilities_1992,shaqfeh_purely_1996, mckinley1996rheological,pakdel1996elastic,pathak_elastic_2004,browne_porescale_2020, steinberg_elastic_2021,datta_perspectives_2022,kumar_transport_2022}. The Reynolds number $Re$ associated with such flows is typically $\ll 1$~\citep{browne_porescale_2020}, and so, this instability is known as a \emph{purely-elastic} instability. Its onset can instead be parameterized using the Weissenberg number, $Wi = N_1/\left(2\sigma\right)$, which compares the strength of elastic stresses, described by the first normal stress difference $N_1$, to viscous stresses, described by the shear stress $\sigma$.

What are the characteristics of elastic instabilities in a porous medium? Despite its apparent simplicity, the answer to this question is unclear. One reason is because the flow is highly sensitive to pore space geometry: studies in idealized one- and two-dimensional (1D and 2D) microfluidics~\citep{anbari_microfluidic_2018,browne_porescale_2020,datta_perspectives_2022} using a variety of polymeric fluids have uncovered a dizzying array of different flow behaviors in different geometries, including symmetry-breaking~\citep{arratia_elastic_2006, dubash_elastic_2012, haward_instabilities_2013,haward_asymmetric_2020,qin_three-dimensional_2020,hopkins_effect_2022}, quasi-periodicity~\citep{zhao_flow_2016,hopkins_upstream_2022}, the formation of pronounced eddies and wakes~\citep{shi_growth_2016,zhao_flow_2016,kawale_elastic_2017,qin_upstream_2019,haward_asymmetric_2020,walkama_disorder_2020,browne_bistability_2020,kumar2021numerical,haward_stagnation_2021,hopkins_upstream_2022,hopkins_effect_2022,dzanic_geometry_2023,chen_influence_2024}, and chaotic dynamics reminiscent of turbulence~\citep{groisman_elastic_2000,groisman_elastic_2004,zilz_geometric_2012,casanellas_stabilizing_2016,qin_characterizing_2017,qin_flow_2019,van_buel_characterizing_2022}. This broad diversity of ways in which the flow instability manifests across different geometries can be challenging to make sense of. As a result, studies probing the unsteady flow in porous media with defined and systematically-tunable geometries are critically needed. Addressing this gap in knowledge is not just fundamentally interesting, but also practically important: there is tremendous interest in harnessing this elastic instability to homogenize fluid flow~\citep{browne_homogenizing_2023}, enhance solute mixing and dispersion~\citep{groisman_efficient_2001, burghelea_chaotic_2004,pathak_elastic_2004,scholz_enhanced_2014,jacob_particle_2017,ducloue_secondary_2019,browne_harnessing_2024}, and mobilize trapped immiscible fluids~\citep{clarke_mechanism_2015, clarke_how_2016,de_viscoelastic_2018,shakeri_effect_2021} in porous media. Furthermore, the onset of instability at the pore scale has been linked to a substantial increase in flow resistance at large scales~\citep{arora_experimental_2002,clarke_mechanism_2015, browne_elastic_2021} and thus has major implications for macroscopic operating conditions. However, in all these cases, a straightforward link between imposed flow conditions, pore space geometry, and the characteristics of the flow instability remains lacking.

A key conceptual advance arising from prior 2D studies is the importance of stagnation points---points at which the local fluid velocity is zero that arise as streamlines diverge/converge around solid boundaries. Stagnation points are therefore dominated by locally extensile flow that generates large tensile stresses in the polymeric fluid that control the onset and strength of the elastic instability~\citep{oztekin1997stability,arratia_elastic_2006,poole2007purely,shi_growth_2016,zhao_flow_2016,qin_upstream_2019,haward2019flow,haward_asymmetric_2020,haward_stagnation_2021,hopkins_upstream_2022,hopkins_effect_2022}. The density and location of these points is dictated by boundary geometry. Hence, identifying when and where stagnation points arise in media of different geometries may provide a straightforward way to disentangle the influence of pore space geometry on the unsteady flow of a polymer solution. Though studies in 2D have provided useful insights, pore space topology---and thus, the occurrence of stagnation points---is fundamentally different in a three-dimensional (3D) porous medium~\citep{lester_chaotic_2016, marafini_suitability_2020}. Therefore, we ask: Where do stagnation points arise in 3D porous media of different geometries, and what are the implications for the unsteady flow of a polymeric fluid? 

\begin{figure*}
\centerline{\includegraphics{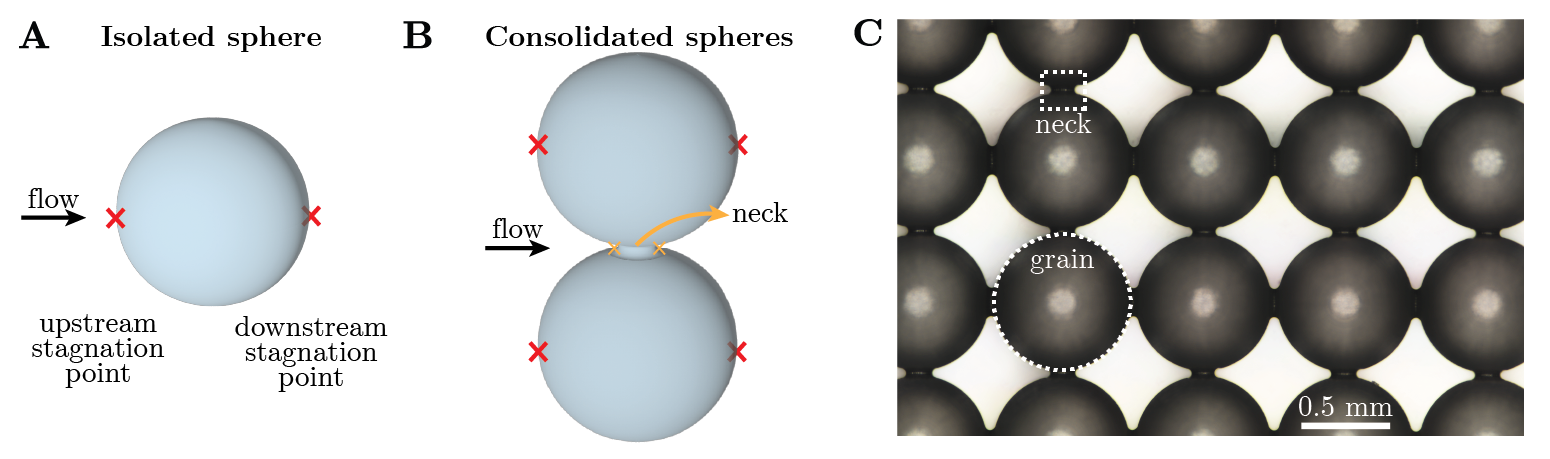}}
  \caption{Stagnation points in flow past spheres. (a) Flow past an isolated sphere with polar stagnation points marked by red $\times$'s. (b) Flow past consolidated spheres joined together by a neck, producing new stagnation points marked by yellow $\times$'s. (c) Image of a consolidated simple cubic packing, taken from top-down in air.}
  \label{schematics}
\end{figure*}

Here, we address this question by directly visualizing the pore-scale flow of an elastic polymer solution inside 3D porous media of defined geometries. We use microfabricated simple cubic (SC) and body-centered cuboid (BC) packings of monodisperse glass spherical grains, which circumvent the geometric complexity of disordered media in which the features of the flow instability can differ from pore to pore~\citep{browne_elastic_2021}. Stokes flow around a single spherical grain has two stagnation points: one at the upstream pole where the fluid streamlines diverge around the grain, and the other immediately downstream where the streamlines reconverge (Figure~\ref{schematics}A). These points would then generate strong local extensional flow that can stretch polymers and thereby control the elastic instability~\citep{haward_molecular_2004}. However, in certain ordered 3D packings, which are often conceptualized as having point-like contacts between adjacent grains, these polar stagnation points are screened from the flow due to the presence of the contacting upstream and downstream grains. Rather, in practice, the contacts between adjacent grains are not discrete points, but have extended catenoid-like geometries arising from grains fusing together (Figure~\ref{schematics}B,C) --- as is the case during consolidation and diagenesis, which are processes inherent in most natural and engineered media. We find that stagnation points are generated at these contacts instead, and these new stagnation points are what control the onset and features of the elastic flow instability, which manifest as a transition from steady to unsteady, time-dependent flow. Moreover, by combining our flow visualization with pressure drop measurements, we show that the extensional flow associated with these contact-associated stagnation points causes the macroscopic flow resistance to sharply increase, helping to rationalize previous pressure drop measurements~\citep{haward_viscosity_2003,odell_viscosity_2006}. Our work thus reveals the previously-overlooked, pivotal role of grain contacts in shaping elastic flow instabilities in 3D porous media.

\section{Materials and Methods} \label{sec:materials_methods}
\begin{figure*}
\centerline{\includegraphics{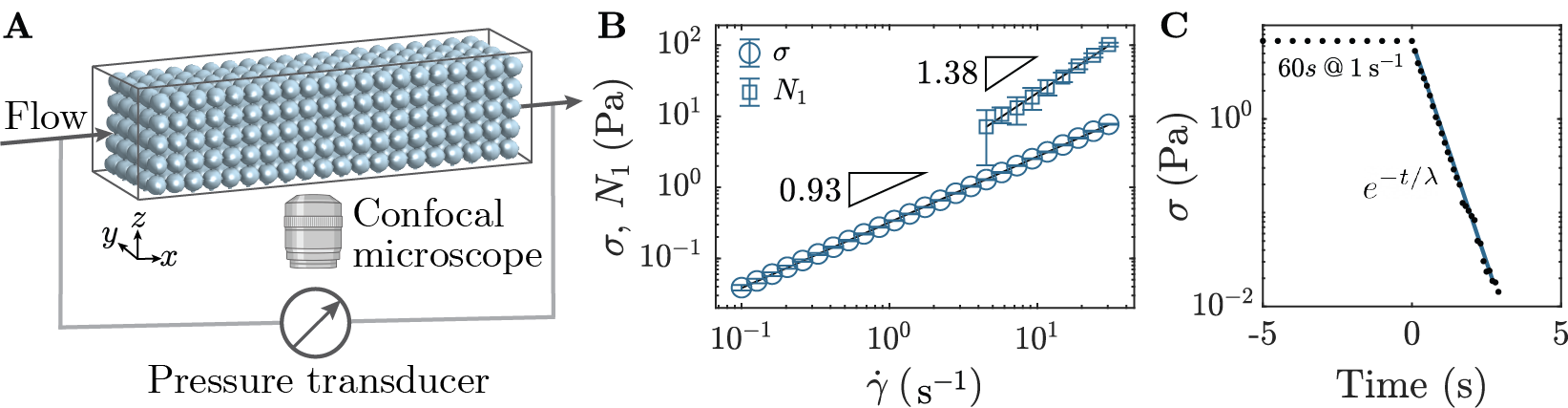}}
  \caption{Experimental set-up and fluid rheology. (A) Schematic of experimental set-up. The microfluidic device containing the microfabricated packing (not to scale) is mounted on the stage of a confocal microscope. A pressure transducer measures the bulk pressure drop across the entire medium. (B) Steady-shear rheology of the 300 ppm HPAM solution. Linear fits correspond to power law models for both the shear stress, $\sigma$, and first normal stress difference, $N_1$. Error bars represent one standard deviation from averaging over 3 replicate flow curves, using new fluid samples for each replicate. In some cases, error bars may be smaller than symbol size. The range of $N_1$ data is limited due to measurement noise of the normal force at lower shear rates. (C) Sample stress relaxation measurement. A controlled shear rate of $1 \: \mathrm{s}^{-1}$ is applied for 60 s and immediately stopped. The longest polymer relaxation time $\lambda$ is found by fitting an exponential stress relaxation, assuming a single relaxation mode.}
  \label{experimental}
\end{figure*}

We investigate flow through two different ordered packings of spherical grains (diameter $D_p = 800 \: \upmu \mathrm{m}$) with an overall square cross-section of area $A_c=16 \: \mathrm{mm^2}$, as schematized in Figure~\ref{experimental}A. The packings are made by selective-laser induced etching of an open-faced slab of fused silica, as detailed previously~\citep{carlson_volumetric_2022}. They have a simple cubic (SC) or body-centered cuboid (BC) unit cell; the latter has a rectangular cuboid unit cell, with its longest edge along the direction of imposed flow, instead of the uniform cubic cell typical of an ideal BC packing, to maintain the same number of unit cells in each cross-section. The SC array contains 5 $\times$ 5 $\times$ 47 packed grains, giving a total medium length of $\Delta L = 3.84$ cm and porosity $\phi = 0.48$. The BC array contains 5 $\times$ 5 $\times$ 23 grains (excluding the sphere in the center of each unit cell), giving a total medium length of $\Delta L= 2.74$ cm and porosity $\phi=0.28$. Importantly, the contacts between adjacent grains are not discrete points; instead, due to the microfabrication process, they are small ``necks" of diameter $D_{\mathrm{neck}} = 170 \: \upmu \mathrm{m}$, measured from confocal imaging, that are representative of granular contacts in many natural and engineered media. To enable polymeric fluid flow through the pore space, we glue flexible Tygon tubing and attach plastic connectors (McMaster-Carr) to each microfluidic device. 

While such 3D media are typically opaque, we build on our prior work and formulate a polymeric fluid whose refractive index is matched to that of the silica; thus, each medium becomes transparent when saturated with the fluid, enabling us to visualize the motion of fluorescent tracers inside the 3D pore space and thereby map the flow field in situ using particle image velocimetry (PIV)~\citep{ datta_spatial_2013,browne_elastic_2021,carlson_volumetric_2022}. We prepare the polymeric fluid by first dissolving partially hydrolyzed polyacrylamide (HPAM; 18 MDa; 30\% carboxylated monomers; Polysciences) in ultrapure Millipore water, and adding glycerol (Sigma-Aldrich), dimethyl sulfoxide (DMSO; Sigma-Aldrich), and sodium chloride (NaCl; Sigma-Aldrich). The final solution composition contains 300 ppm HPAM, 82.6 wt\% glycerol, 10.4 wt\% DMSO, and 1 wt\% NaCl. This solution falls in the dilute concentration regime, $c/c^* = 0.5$, where $c^* = 600 \pm 300$ ppm is the polymer overlap concentration~\citep{browne_elastic_2021}.

To prepare each porous medium before performing flow visualization, we sequentially flush the device with milliQ water, isopropanol (IPA; Fisher Chemical), and milliQ water again, before filling the medium with glycerol. We then displace the glycerol with the polymer solution at a constant flow rate $Q = 3 \: \mathrm{mL/hr}$ using a Harvard Apparatus PHD 2000 syringe pump for at least 3 hours to pre-saturate the medium with the polymer solution. During each flow visualization experiment, we apply a ramp of increasing flow rate, where we apply flow for at least 90 minutes at each new flow rate before beginning imaging to achieve a equilibrated flow state. To visualize the flow, we seed the fluid with 7 ppm of fluorescent 1 $\upmu$m carboxylate-modified polystyrene tracer particles (Invitrogen), which have excitation at 480-510 nm and emission at 515 nm. Particles are excited by a 488 nm laser and detected by a 500-550 nm sensor. For imaging, the medium is placed on the stage of a Nikon A1R+ laser scanning confocal fluorescence microscope. We use a 10$\times$ objective to interrogate a 1272.79 $\upmu$m $\times$ 1272.79 $\upmu$m field of view at 512 $\times$ 512 pixels (2.46 $\upmu$m per pixel). The optical sectioning provided by the confocal microscope and fluid refractive index matching enables clear imaging of a $z$-slice of the flow field within the 3D medium. We use the confocal resonant scanner to capture videos at 30 fps with an optical section thickness of 7.39 $\upmu$m at depths $400 < z < 1200 \: \upmu$m from the bottom of the medium. We image 10 randomly selected pores across the porous medium and do not observe significant differences in flow behavior for pores beyond $\sim 1-2$ sphere layers near the inlet, presumably due to the identical geometry of each pore throughout the packing. Thus, we select the pore located in the center of the medium for a majority of the subsequent analysis presented in the results.

We use a built-in MATLAB particle image velocimetry toolbox (PIVlab) to obtain 2D temporally-resolved velocity fields~\citep{thielicke_pivlab_2014}. We use the FFT window deformation PIV algorithm with 4 passes of decreasing interrogation area (64-32-16-16 pixels) and 50\% overlap to obtain velocity fields, $\mathbf{u}=(u,v)$, on a spatial grid, $\mathbf{x}=(x,y)$, with spatial resolution of $\Delta x = \Delta y = 20.2 \: \upmu$m and a temporal resolution of 33 ms. We use image processing software (ImageJ) to generate fluid pathline videos and time-averaged pathline images by applying a moving average of mean pixel intensity over a selected number of frames. 

To characterize the macroscopic flow resistance, we measure the pressure drop across the entire medium by connecting a differential pressure transducer (0-30 psi, Omega PX26) in parallel to the device. We apply a constant flow rate using a syringe pump and wait 60 minutes before recording the pressure signal while sampling at a rate of 10 Hz for 90 minutes at each applied flow rate. We calculate a time-averaged pressure signal, adjusted for measurement noise by normalizing by the average pressure signal in the absence of flow. We then determine an apparent viscosity from the pressure drop measurements using Darcy's law: $\eta_{\mathrm{app}}=\Delta Pk/\left((Q/A_c)\Delta L\right)$, where $k$ is the absolute permeability of the porous medium. We determine the permeability using a flow rate ramp of pure water in each medium, extracting $k$ from the slope of the linear fit to the pressure, $\Delta P$, versus flow rate, $Q$. The permeabilities of each medium are $k_{\mathrm{SC}}=567 \pm 9 \: \upmu \mathrm{m}^2$ and $k_{\mathrm{BC}}=197 \pm 2 \: \upmu \mathrm{m}^2$.

We characterize the bulk solution rheology using an Anton-Paar MCR501 rheometer fitted with a truncated cone-plate measuring tool (CP50-2: 50 mm, 2$^\circ$, 53 $\upmu \mathrm{m}$ gap) operating at a controlled temperature of 25$^\circ$C. Figure~\ref{experimental}B presents steady-shear rheology measurements of the shear stress ($\sigma$) and first normal stress difference ($N_1$) over shear rates $0.1 < \dot{\gamma} < 50 \ \mathrm{s^{-1}}$. As shown by the circles, the shear stress increases approximately linearly with shear rate; a power law fit ($\sigma \sim \dot{\gamma}^{n_{\sigma}}$) has an exponent $n_{\sigma} = 0.93$, indicating that the polymer solution is very weakly shear-thinning, so we do not expect effects of shear-thinning or shear-banding to influence the flow behavior. We use the shear stress at the lowest measured shear rate ($\dot{\gamma}=0.1 \: \mathrm{s^{-1}}$) to estimate the zero-shear viscosity as $\eta_0 \approx \sigma/\dot{\gamma}|_{0.1 \: \mathrm{s^{-1}}}= 0.4 \: \mathrm{Pa \: s}$. The first normal stress difference also increases with shear rate as a power law ($N_1 \sim \dot{\gamma}^{n_{N_1}}$) with exponent $n_{N_1} = 1.38$, indicating that the polymer solution is elastic. We also use shear stress relaxation measurements to estimate the solution relaxation time $\lambda$~\citep{liu_longest_2007,liu_concentration_2009}; an example is shown in figure~\ref{experimental}C. Averaging over 3 replicates and fitting the data to a single-mode exponential decay ($\sigma \sim e^{-t/\lambda}$) yields a longest relaxation time of $\lambda = 0.4 \pm 0.1$~s.

\section{Results}
\label{sec:results}

\subsection{Pore-scale imaging reveals the onset of an elastic instability above a threshold $Wi$}
First, we use in situ visualization to probe how the pore-scale polymer solution flow changes with increasing flow rate in both ordered packings. The characteristic interstitial shear rate is given by $\dot\gamma_I=U_I/\sqrt{k/\phi}$, where $U_I=Q/\left(\phi A_c\right)$ and $\sqrt{k/\phi}$ are the characteristic interstitial flow speed and pore length scale, respectively~\citep{zami-pierre_transition_2016,berg_shear_2017}. We estimate the Reynolds number as $Re=\rho U_I D_p/\eta_I$, where $\rho$ is the fluid density and $\eta_I$ is the bulk shear viscosity evaluated at the interstitial shear rate $\dot\gamma_I$. We also compute the Weissenberg number, $Wi=N_1/(2\sigma)$, directly from the bulk shear measurements of $N_1$ and $\sigma$ evaluated at $\dot\gamma_I$ (Figure~\ref{experimental}B). Our experiments probe $10^{-5} < Re < 10^{-3} \ll 1$ and $1 < Wi < 5$.

\begin{figure*}
  \centerline{\includegraphics{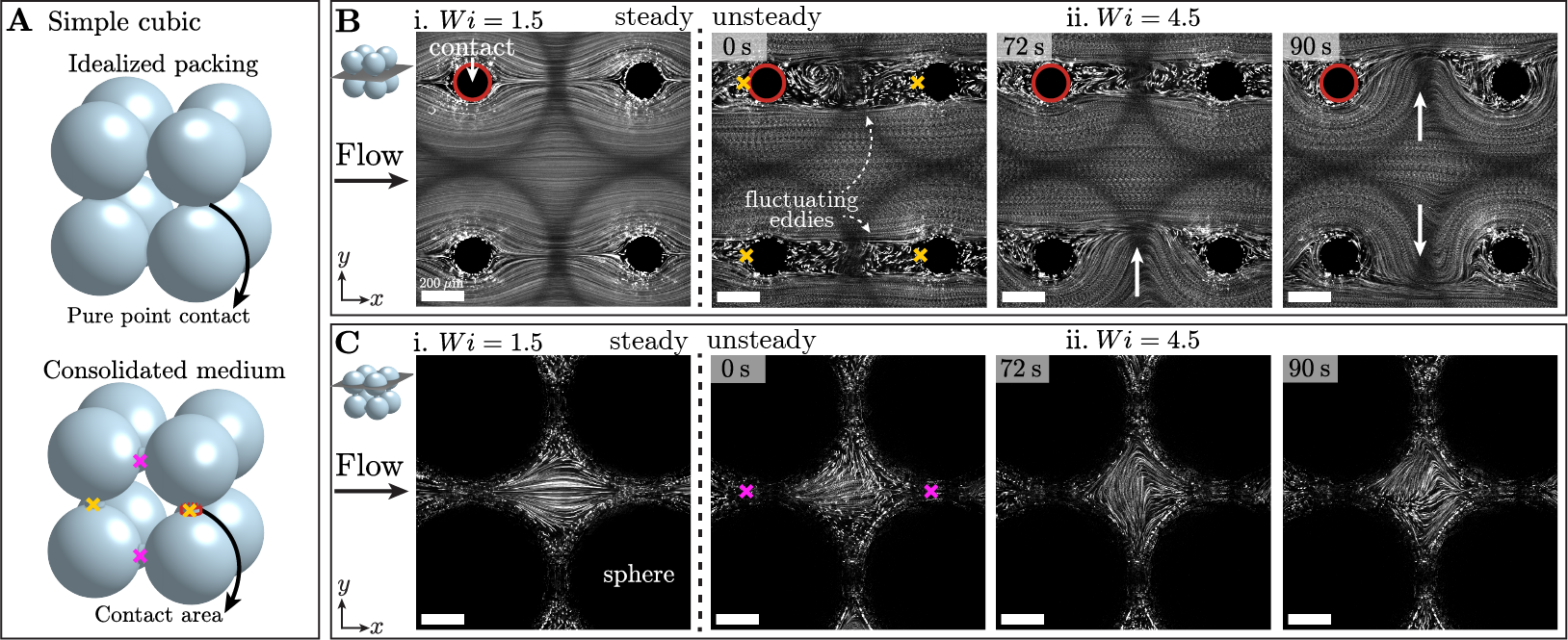}}
  \caption{Elastic instability in SC packing: 3D fluctuating eddies. (A) Schematics contrasting an idealized SC unit cell (top) containing pure point contacts with the consolidated SC unit cell (bottom) used in this work. The small region outlined in red between spheres marks an example of a neck-like inter-grain contact. Yellow $\times$ demarcate stagnation points arising at the contacts. Pink $\times$ also mark contact-generated stagnation points, but for contacts that are rotated 90$^\circ$ with respect to the $z$-plane shown in the pathline images. (B) Pathline images corresponding to the $z$-plane containing the inter-grain contacts (example contact circled in red). i. Steady flow field at $Wi = 1.5$. Pathlines are generated by averaging successive frames over 20 second intervals. Scale bar is 200 $\upmu$m. ii. Unsteady flow field at $Wi = 4.5$ for 3 different time points. Highly fluctuating eddies develop between adjacent contacts. Pathlines are generated by averaging successive frames over 2 second intervals. White arrows indicate cross-flow ``burst" events between adjacent pores. (C) Pathline images in the $z$-plane located 1/2-sphere diameter above the $z$-plane in (B). i. Steady flow field at $Wi = 1.5$. ii. Unsteady flow field at $Wi = 4.5$ for 3 different time points, where the flow field fluctuates chaotically.}
  \label{sc_streakline}
\end{figure*}

\subsubsection{Simple cubic packing: 3D fluctuating eddies}
The idealized representation of the SC unit cell assumes point contacts between adjacent grains, as schematized in the top panel of Figure~\ref{sc_streakline}A. By contrast, real-world media typically have spatially-extended contacts between grains due to consolidation or diagenesis, in which grains fuse together due to the weight of overlying material or biogeochemical processes, respectively; indeed, a 3D packing with purely point contacts would not sustain rigidity under flow~\citep{deresiewicz1958mechanics,biot1963theory,winkler1983contact,surdam1989organic,mueth1998force,makse2004granular,burley2009sandstone,andreotti2013granular,horabik2016parameters}. Our microfabrication process recreates this inherent feature of granular packings. In particular, the unit cell of our microfabricated SC packing has neck-like contacts between adjacent grains---as shown by the red outlines in the schematic of Figure~\ref{sc_streakline}A (lower panel) and confocal micrographs in Figure~\ref{sc_streakline}B. As we detail below, this simple geometric feature has pivotal implications for the unsteady flow behavior.

At low $Wi = 1.5$, the flow in the interstitial pore space is steady, as shown by the steady fluid pathlines shown in Figure~\ref{sc_streakline}B i. By contrast, for large enough $Wi$, an elastic instability is generated by the additional stagnation points directly upstream and downstream of the inter-grain contacts (yellow $\times$ in Figure~\ref{sc_streakline}A--B). Eddies form on the upstream faces of the contacts and grow in length with increasing $Wi$, localizing in the regions between neighboring contacts, as shown in SI Video 1 for a range of $1.5<Wi<4.5$. Above a threshold $Wi=Wi_c\approx3$, these eddies become highly unsteady, as exemplified by the three different panels in Figure~\ref{sc_streakline}B ii for the case of $Wi = 4.5$. They exhibit large fluctuations in size due to ``bursts" of transverse cross-flow between pores, as indicated by the white arrows, corroborating the previous observations of~\cite{carlson_volumetric_2022}.

The variation of this flow field in 3D can be seen by imaging other $z$ planes of the SC unit cell. For example, Figure~\ref{sc_streakline}C shows additional pathline images of the flow for a $z$-plane located half a grain diameter above that shown in Figure~\ref{sc_streakline}B; imaging this plane is equivalent to imaging \textit{through} an eddy due to the symmetry of the packing. Stagnation points generated at inter-grain contacts rotated 90$^\circ$ relative to the plane of imaging are denoted by the pink $\times$. Again, at sufficiently large $Wi$, the eddies are highly unsteady, as shown by the crossing pathlines and distinct changes in mean flow direction at different time points in Figure~\ref{sc_streakline}C ii. SI Videos 2 and 3 show the flow behavior at $Wi=1.5$ (steady) and $Wi = 4.5$ (unsteady) for additional $z$-planes spaced approximately 50 $\upmu$m apart that span half of the characteristic unit cell volume.

\begin{figure*}
  \centerline{\includegraphics{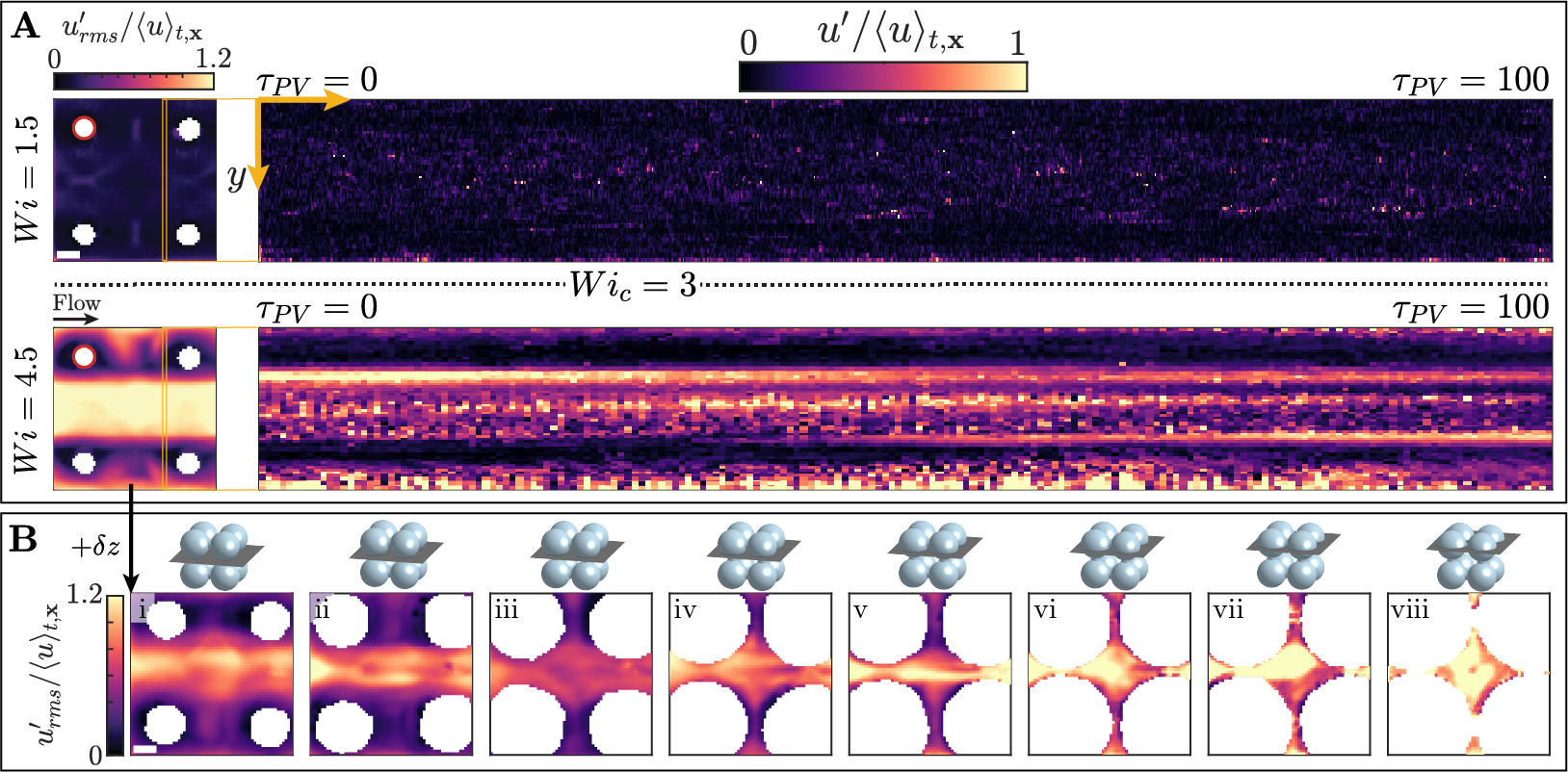}}
  \caption{Spatiotemporal features of the SC eddy instability. (A) Left: Normalized r.m.s. velocity fluctuation maps for steady ($Wi=1.5$, top left) and unsteady ($Wi=4.5$, bottom left) flow states. Scale bar is $200 \: \upmu$m. Red circles mark examples of inter-grain contacts. Right: Kymographs of the normalized velocity fluctuations corresponding to the column of pixels boxed in the colormaps on the left. Time $t$ is reported as the number of injected pore volumes, $\tau_{PV}=t/\left(\mathrm{PV}/Q\right)$, where one pore volume (PV) is defined as the void volume of one unit cell of the packing. (B) Normalized r.m.s. velocity fluctuation maps for additional $z$-planes at $Wi=4.5$ spanning half of the unit cell. Sequential $z$-planes are spaced approximately $50 \: \upmu$m apart.}
  \label{sc_spatiotemporal}
\end{figure*}

To more clearly represent these complex dynamics, we map the normalized root-mean-square (r.m.s.) velocity fluctuations, $u_{\mathrm{rms}}'/\langle u \rangle_{t,\mathrm{\mathbf{x}}}$, where $\langle u \rangle_{t,\mathrm{\mathbf{x}}}$ is the mean velocity magnitude averaged over time $t$ and space $\mathbf{x}$. The fluctuations corresponding to the plane imaged in Figure~\ref{sc_streakline}B are shown in Figure~\ref{sc_spatiotemporal}A. At $Wi = 1.5$, the flow is steady and thus no appreciable fluctuations are seen. Above $Wi=Wi_c\approx3$, sustained flow fluctuations larger than the PIV noise threshold ($u'/\langle |\mathbf{u}| \rangle_t > 0.2$) start to appear. At $Wi = 4.5$, velocity fluctuations are largest in the bulk void space of the unit cell, reflecting the cross-flow events that intermittently wash away eddies. Figure~\ref{sc_spatiotemporal}A also presents kymographs (showing space and time on the vertical and horizontal axes, respectively) of the normalized instantaneous velocity fluctuations for a selected column of pixels located in a region upstream of the inter-grain contacts over a duration $\tau_{PV}=100$. The kymograph for $Wi = 1.5$ shows no fluctuations over space and time because the flow is steady, whereas the kymograph for $Wi=4.5$ shows velocity fluctuations persisting over both space and time. The dark purple bands in the $Wi = 4.5$ kymograph correspond to the eddy-dominated regions, and eddy-washing events are observed as a thinning of these dark purple bands over time. Figure~\ref{sc_spatiotemporal}B presents maps of the normalized r.m.s. fluctuations for additional $z$-planes (panels i-viii) spanning half of the characteristic unit cell volume for $Wi = 4.5$. The largest velocity fluctuations are located in the bulk void space and the $z$-plane in panel viii. These $z$-planes correspond to regions of the unit cell containing contacts that are directly accessible to the impinging flow. Note that the neck-like contacts in Figure~\ref{sc_spatiotemporal}B viii are rotated $90^\circ$ with respect to the contacts in the $z$-plane shown in Figure~\ref{sc_spatiotemporal}A; although these contacts are rotated with respect to the imaging plane, they generate equivalent stagnation points to those in the $z$-plane presented in Figure~\ref{sc_spatiotemporal}A. Altogether, our flow visualization demonstrates that the additional stagnation points generated by inter-grain contacts causes the elastic instability to develop locally near the contacts compared to other regions in the pore volume.

\begin{figure*}  \centerline{\includegraphics{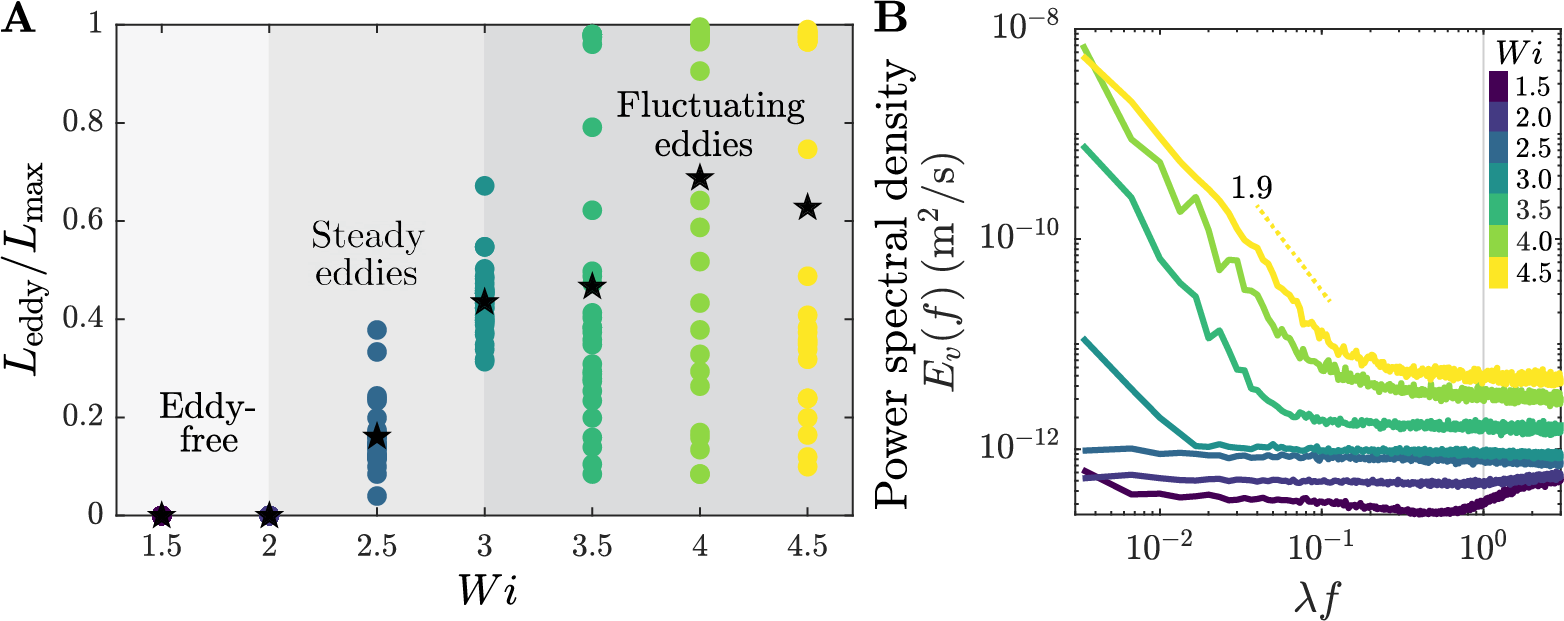}}
  \caption{Characterization of eddy development and spectral analysis of velocity fluctuations in the SC packing. (A) Quantification of eddy regimes: eddy length normalized by the maximum eddy length. For $Wi < 2$, no eddies are present and the flow is steady. For $2 < Wi < 3$, the average eddy length increases with $Wi$ and remains fairly steady over time. Above $Wi = 3$, eddy lengths can reach their maximum, but are also highly fluctuating as reflected by the large spread in the length distributions. Distributions of eddy lengths at each Wi correspond to measurements sampled every 10 seconds over 5 minutes of imaging. Black stars denote the temporally-averaged eddy length. (B) Temporal power spectral densities $E_v(f)$ for the transverse velocity fluctuations $v'$ in the SC packing. We determine the temporal power spectrum at each PIV grid location in the field of view, then average the temporal spectra over all positions to reduce noise. The onset of steep power spectral decay coincides with the onset of instability $Wi_c \approx3$. Power laws are fitted to the regions of the spectra with the steepest decay. At $Wi = 4.5$, the decay exponent is $\alpha = 1.9$. The gray vertical line marks the inverse of the longest polymer relaxation time, $1/\lambda$ ($\lambda = 0.4$ s).}
  \label{sc_eddy}
\end{figure*}

We further characterize this flow instability by measuring the normalized eddy length $L_{\mathrm{eddy}}/L_{\mathrm{max}}$ every 10 seconds over 5 minutes of imaging for each $Wi$ tested. We find three distinct flow regimes, as shown in Figure~\ref{sc_eddy}A. For $Wi < 2$, the flow is steady and eddy-free. At larger $2 < Wi < 3$, eddies begin to form on the upstream faces of sphere contacts and grow increasingly longer with $Wi$ (SI video 1). In this regime, the eddies remain fairly steady in length, reflected by a narrower spread of the sampled eddy lengths. Above the threshold $Wi=Wi_c\approx3$, the eddies are highly unsteady, with lengths that fluctuate drastically between nearly zero to as large as the distance between adjacent contacts. These fluctuations are also reflected in the power spectral density $E_v(f)=\lvert \mathrm{FFT}(Y)\rvert^2/\left(L F_s\right)$ of transverse velocity fluctuations $v'$ of different temporal frequencies $f$; here, $\mathrm{FFT}(Y)$ is the fast Fourier transform of the temporal velocity fluctuation signal $Y$, $L$ is the length of the signal, and $F_s$ is the signal sampling rate. As shown by the green to yellow curves in Figure~\ref{sc_eddy}B, above the threshold $Wi_c \approx 3$, the spectra begin to exhibit a power law decay $E_v(f)\sim f^{-\alpha}$ at frequencies smaller than $1/\lambda$. The total power contained in these  fluctuations, given by the area under the curve of $E_v(f)$, increases with $Wi$, with most of the energy contained in frequencies smaller than $1/\lambda$: the flow fluctuations occur over a broad range of time scales primarily longer than the longest polymer relaxation time. The characteristic power law exponent $\alpha \approx 2$ is consistent with previous measurements in model 3D porous media at similar $Wi$~\citep{browne_elastic_2021,carlson_volumetric_2022}. This exponent is lower than the value of $\alpha \approx 3.5$ that is typically referenced as a spectral signature of purely ``elastic turbulence'' ~\citep{groisman_elastic_2000,fouxon_spectra_2003,groisman_elastic_2004,steinberg_scaling_2019,varshney_elastic_2019,van_buel_characterizing_2022}, suggesting that our experiments probe a transitional unsteady flow state en route to fully-developed elastic turbulence.

\begin{figure*}
  \centerline{\includegraphics{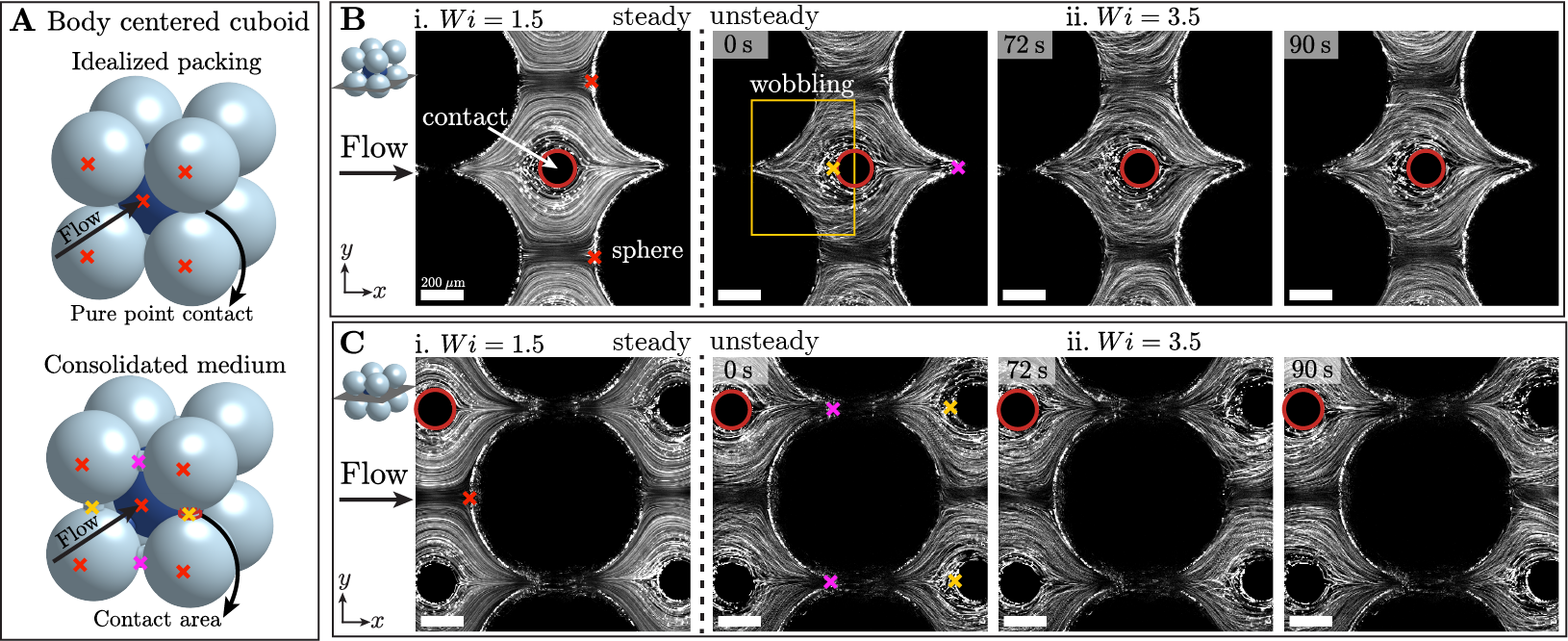}}
  \caption{Elastic instability in the BC packing: chaotic, ``wobbling" pathlines. (A) Schematics contrasting an idealized BC cuboid unit cell (top) containing pure point contacts with the consolidated BC cuboid unit cell (bottom) used in this work. The small neck-like region outlined in red between spheres marks an example of an inter-grain contact. Red $\times$ correspond to stagnation points generated at sphere surfaces. Yellow $\times$ correspond to stagnation points arising at the contacts. Pink $\times$ also mark contact-generated stagnation points, but for contacts that are rotated 90$^\circ$ with respect to the $z$-plane shown in the pathline images. (B) Pathline images corresponding to the $z$-plane containing the inter-grain contacts (circled in red). i. Steady flow field at $Wi = 1.5$. Pathlines are generated by averaging frames over 20 seconds. Scale bar is 200 $\upmu$m. ii. Unsteady flow field at $Wi = 3.5$ for 3 different time points. The instability is characterized by crossing pathlines; the yellow box highlights a region of pronounced pathline wobbling upstream of a contact. Pathlines are generated by averaging frames over 2 seconds. (C) Pathline images in the $z$-plane located 1/2 sphere diameter above the $z$-plane in (B). i. Steady flow field at $Wi = 1.5$. ii. Unsteady flow field at $Wi = 3.5$ for 3 different time points. The pore space geometry in this $z$-plane is equivalent to translating the field-of-view for the $z$-plane in (B) due to the symmetry of the BC unit cell.}
  \label{bc_streakline}
\end{figure*}

\subsubsection{Body-centered cuboid packing: wobbling pathlines}
We next consider flow of the same polymer solution through the BC packing. This medium also has neck-like contacts between adjacent grains (yellow and pink $\times$ in Figure~\ref{bc_streakline}A). As we detail below, by generating stagnation points, these contacts shape the unsteady flow behavior---just as in the SC case. However, the BC packing contains another, distinct set of stagnation points: those at the polar upstream/downstream grain surfaces (red $\times$ in Figure~\ref{bc_streakline}A), as in the case of Stokes flow around a single sphere. As we show below, this change in geometry changes features of how the elastic instability manifests in BC packings.

Figure~\ref{bc_streakline}B shows pathline images of the flow in a $z$-plane containing inter-grain contacts. At low $Wi = 1.5$, the flow is steady with steady, uncrossing pathlines (panel i). As $Wi$ increases above a threshold value, the elastic instability develops upstream of the contacts (panel ii)---just as in the SC case---rather than at the polar upstream/downstream stagnation points. However, this unsteady flow is different from that in the SC case in two ways. First, the threshold $Wi_c\approx2$ is slightly smaller than the threshold $Wi_c\approx3$ in the SC packing. Second, in the BC packing, the elastic instability does not manifest as unsteady eddies, but rather pronounced pathline ``wobbling". An example is shown by the crossing pathlines upstream of the contacts shown in the boxed region in  Figure~\ref{bc_streakline}B ii for $Wi=3.5$. SI video 4 shows this transition from steady to unsteady flow, with increasingly chaotic pathline wobbling with increasing $Wi$. The instability is initially generated upstream of the inter-grain contacts before fluctuations start to persist throughout the entire pore space at $Wi \approx 4$.

Figure~\ref{bc_streakline}C shows pathline images for the $z$-plane located half a grain diameter above that shown in Figure~\ref{bc_streakline}B. We observe similar wobbling behavior upstream of the contacts, which reflects the fact that these two $z$-planes are symmetry planes due to the periodicity of the BC packing. SI videos 5 and 6 provide additional pathline imaging for steady ($Wi=1.5$) and unsteady ($Wi = 3.5$) flow states across 9 intermediate $z$-planes that span half of the characteristic unit cell volume. At sufficiently high $Wi$ above the onset of the instability ($Wi_c\approx2$), the flow throughout the entire pore volume is unsteady, as shown in SI video 6.

\begin{figure*}
\centerline{\includegraphics{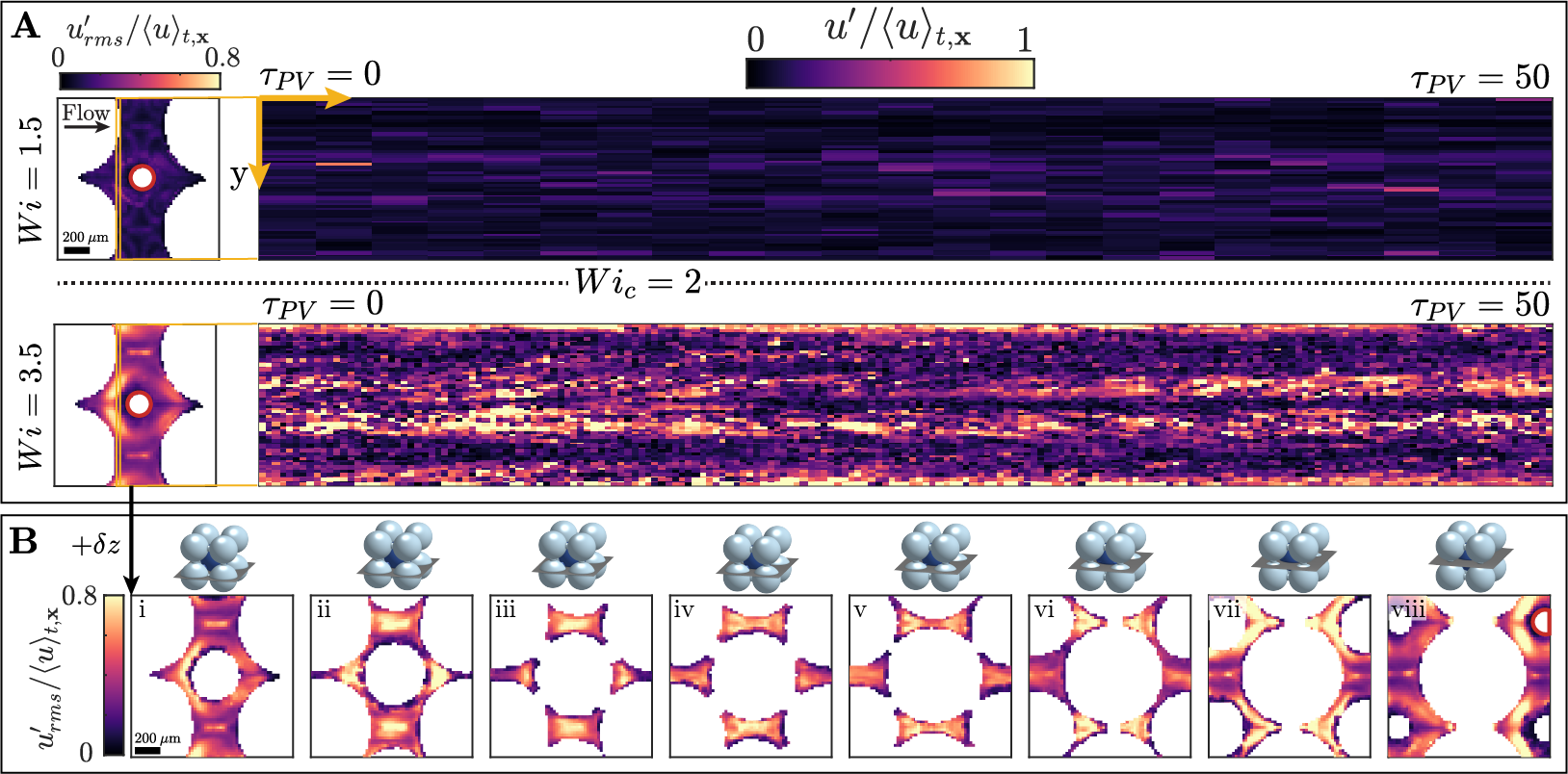}}
  \caption{Spatiotemporal features of the BC wobbling instability. (A) Left: Normalized r.m.s. velocity fluctuation colormaps for steady ($Wi=1.5$, top left) and unsteady ($Wi=3.5$, bottom left) flow states. Right: Kymographs of the normalized velocity fluctuations corresponding to the column of pixels boxed in colormaps to the left. For $Wi = 1.5$, each column in the kymograph corresponds to a chunked average over 15 frames to reduce noise. No averaging is applied to the $Wi = 3.5$ kymograph. Time $t$ is reported as the number of injected pore volumes, $\tau_{PV}=t/\left(\mathrm{PV}/Q\right)$, where one pore volume (PV) is defined as the void volume of one unit cell of the packing. (B) Normalized r.m.s. velocity fluctuation maps for additional $z$-planes at $Wi=3.5$ spanning half of the unit cell. Sequential $z$-planes are spaced approximately $50 \: \upmu$m apart.}
  \label{bc_spatiotemporal}
\end{figure*}

\begin{figure}
  \centerline{\includegraphics{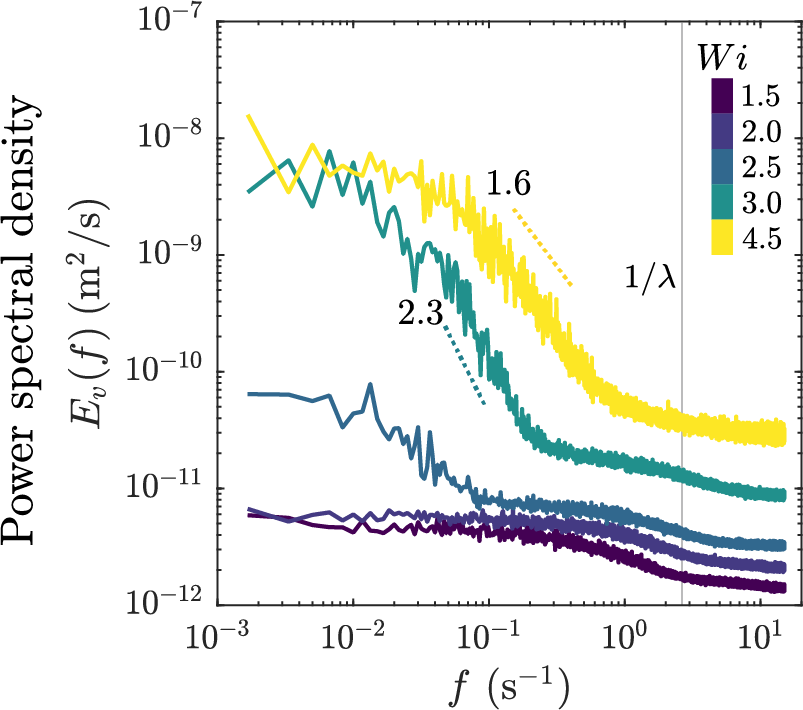}}
  \caption{Temporal power spectral densities $E_v(f)$ for the transverse velocity fluctuations $v'$ in the BC packing at various $Wi$. The onset of power spectral decay coincides with the onset of instability, $Wi_c \approx 2$. Power laws are fitted to the regions of the spectra with the steepest decay. The gray vertical line marks the inverse of the longest polymer relaxation time ($\lambda = 0.4$ s).}
  \label{bc_psd}
\end{figure}

We again map the r.m.s. velocity fluctuations normalized by the mean flow speed, $u_{\mathrm{rms}}'/\langle u \rangle_{t,\mathrm{\mathbf{x}}}$, as shown in Figure~\ref{bc_spatiotemporal}A. At $Wi = 1.5$, the flow is steady with no noticeable fluctuations. At $Wi = 3.5$ above the onset of instability, pronounced velocity fluctuations occur upstream of the inter-grain contact, concomitant with the unsteady wobbling behavior seen in the pathline imaging (Figures~\ref{bc_streakline}B and C). Figure~\ref{bc_spatiotemporal}A also presents kymographs of the normalized instantaneous velocity fluctuations $u'/\langle u \rangle_{t,\mathrm{\mathbf{x}}}$ over 50 injected pore volumes for a column of pixels located upstream of the contact area. No spatiotemporal fluctuations occur at $Wi = 1.5$, contrasted by bursts of large fluctuations at $Wi = 3.5$ that display no characteristic regularity over time; consequently, the flow does not exhibit a periodic state during the transition to instability. Figure~\ref{bc_spatiotemporal}B provides maps of the normalized r.m.s. velocity fluctuations for unsteady flow at $Wi=3.5$ across additional $z$-planes spanning half of the characteristic unit cell volume. Fluctuations in the planes containing the contacts (Figures~\ref{bc_spatiotemporal}A and B viii) are largest upstream of the contacts, while fluctuations in the intermediate planes (Figures~\ref{bc_spatiotemporal}B ii-vii) are less localized throughout the pore space, perhaps due to the more uniform curvature of the geometry.

Figure~\ref{bc_psd} presents the temporal power spectral densities $E_v(f)$ of the transverse velocity fluctuations $v'$ in the BC packing at various $Wi$. As in the SC case, the fluctuations are persistent: most of the kinetic energy in the velocity fluctuations is contained in frequencies smaller than $1/\lambda$. Also, the total power contained in these fluctuations increases with $Wi$. Finally, the 
spectra similarly exhibit a broad power law decay $E_v(f)\sim f^{-\alpha}$, with $\alpha \approx 2$, above the threshold $Wi_c = 2$---just as in the SC packings, and consistent with previous measurements at similar $Wi$ in 3D porous media~\citep{browne_elastic_2021,carlson_volumetric_2022}. As in the SC case, this exponent suggests that our experiments probe a transitional state that arises en route to fully-developed purely elastic turbulence.

\subsection{Local Pakdel-McKinley criterion rationalizes the role of inter-grain contacts in the onset of the elastic instability}
Our experiments have demonstrated that, in both the SC and BC geometries, the spatially-extended contacts formed between adjacent grains generate stagnation points that lead to the development of the elastic instability. For the BC packing, which also has stagnation points at the upstream/downstream poles of grains, the points associated with grain contacts instead generate the instability. As we show next, these findings can be rationalized within the conventional theoretical framework that has been used to understand the onset of elastic instabilities in other, simpler geometries.

In particular, to account for the role of both streamline curvature and fluid elasticity in destabilizing the flow,~\cite{mckinley_rheological_1996} proposed a dimensionless parameter, the $M$ parameter, that aims to describe the onset of elastic instabilities for different solution rheology and flow geometries. The $M$ parameter is defined as $M=\sqrt{\frac{\lambda U}{\mathcal{R}}\frac{N_1}{\eta_0 \dot{\gamma}}}$, where $U$ is the characteristic flow speed and $\mathcal{R}$ is the streamline radius of curvature; it thus requires both sufficient streamline curvature and elastic normal stresses for the instability to occur. Critical values of $M_c \approx 1-10$ have been reported for a range of model geometries~\citep{mckinley_rheological_1996, pakdel_elastic_1996}, although a universal transition value of $M$ has not been found experimentally~\citep{casanellas_stabilizing_2016}. To assess whether this parameter can help describe the onset of the elastic flow instability in our experiments, we use the flow fields measured in both packings to directly compute $M=\sqrt{(N_1(\dot{\gamma})/\eta_0 \dot{\gamma}) \: (\lambda |\mathrm{u}|\kappa)}$, where $\dot{\gamma}$ is the magnitude of the rate-of-strain tensor $|\mathbf{D}|=\sqrt{2\mathbf{D}:\mathbf{D}}$, $\eta_0$ is the fluid zero-shear viscosity, $|\mathrm{u}|$ is the local velocity magnitude, and $\kappa$ is the local streamline curvature, which we determine as a 2D local streamline curvature given that the experimentally-acquired flow fields are in 2D. 

\begin{figure*}
  \centerline{\includegraphics{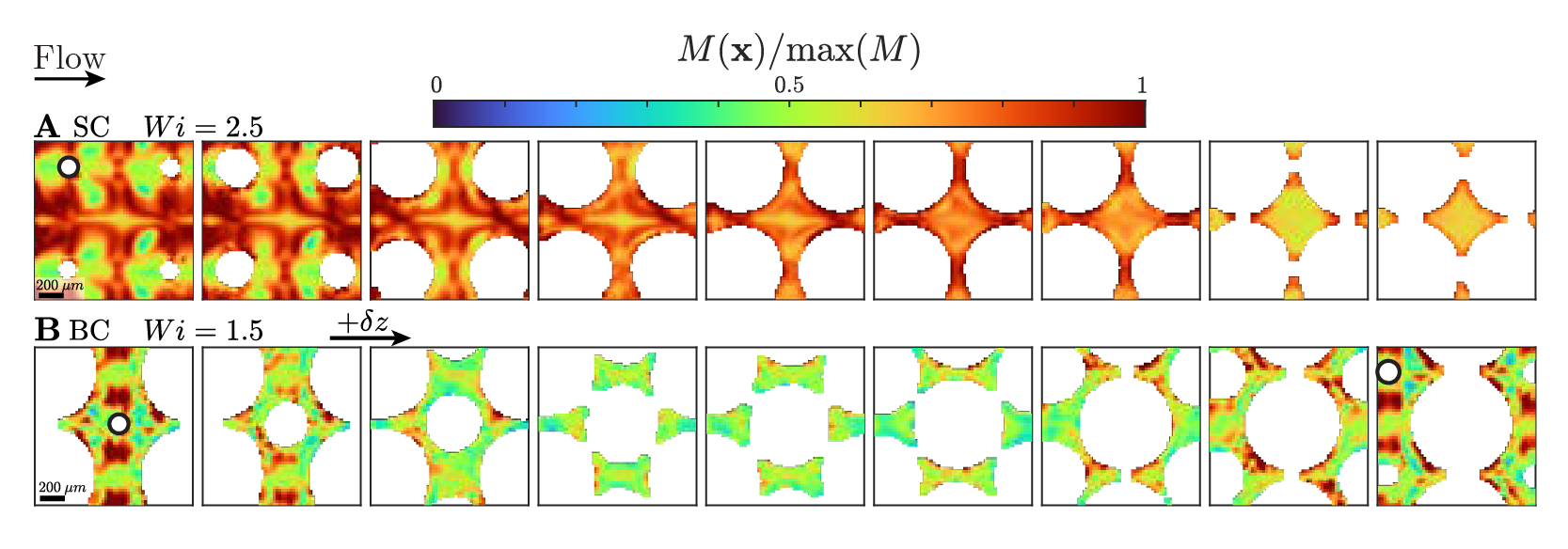}}
  \caption{Colormaps of the local $M$ parameter normalized by the maximum $M$ value for the (A) SC ($Wi = 2.5$) and (B) BC ($Wi = 1.5$) packings at flow conditions below the onset of instability. The largest relative $M$ values for both sphere packings are located in the pore space near the inter-grain contacts, suggesting that the contacts generate sufficient streamline curvature and elastic stress for the instability to develop. The local $M$ fields are temporally averaged over 2 minutes of imaging to reduce noise. Maps are presented for 9 $z$-planes (from left to right) spanning half of the characteristic unit cell, and sequential $z$-planes are spaced approximately $50 \: \upmu$m apart. Example contacts are marked by black circles.The maximum values of M at these flow conditions are $M_{\mathrm{max}}= 1.7$ and 0.82 for the SC and BC packings, respectively.}
  \label{Mparameter}
\end{figure*}

Figure~\ref{Mparameter} shows the resulting colormaps of $M$ normalized by its maximum value, $M(\mathrm{x})/\mathrm{max}(M)$, taken at $Wi$ values just below the instability threshold for 9 different $z$-planes spanning half of the characteristic unit cell volume. For the SC packing (Figure~\ref{Mparameter}A), the regions of highest $M$ are located in the $z$-planes nearest the inter-grain contacts. For the BC packing (Figure~\ref{Mparameter}B), the regions of highest local $M$ occur as lobes on either side of the contacts. In both geometries, the highest streamline curvature under steady flow conditions occurs locally at the contacts, which have a maximum curvature $\kappa\sim 1/R_{\mathrm{neck}}$. These measurements demonstrate that the flow around inter-grain contacts---rather than around the upstream/downstream poles traditionally considered for isolated spheres---has the largest streamline curvature, and therefore, these contacts control the onset of the elastic instability as described theoretically by ~\cite{mckinley_rheological_1996}.

\subsubsection{Transition to instability collapses across different porous medium geometries}
\begin{figure*}
\centerline{\includegraphics{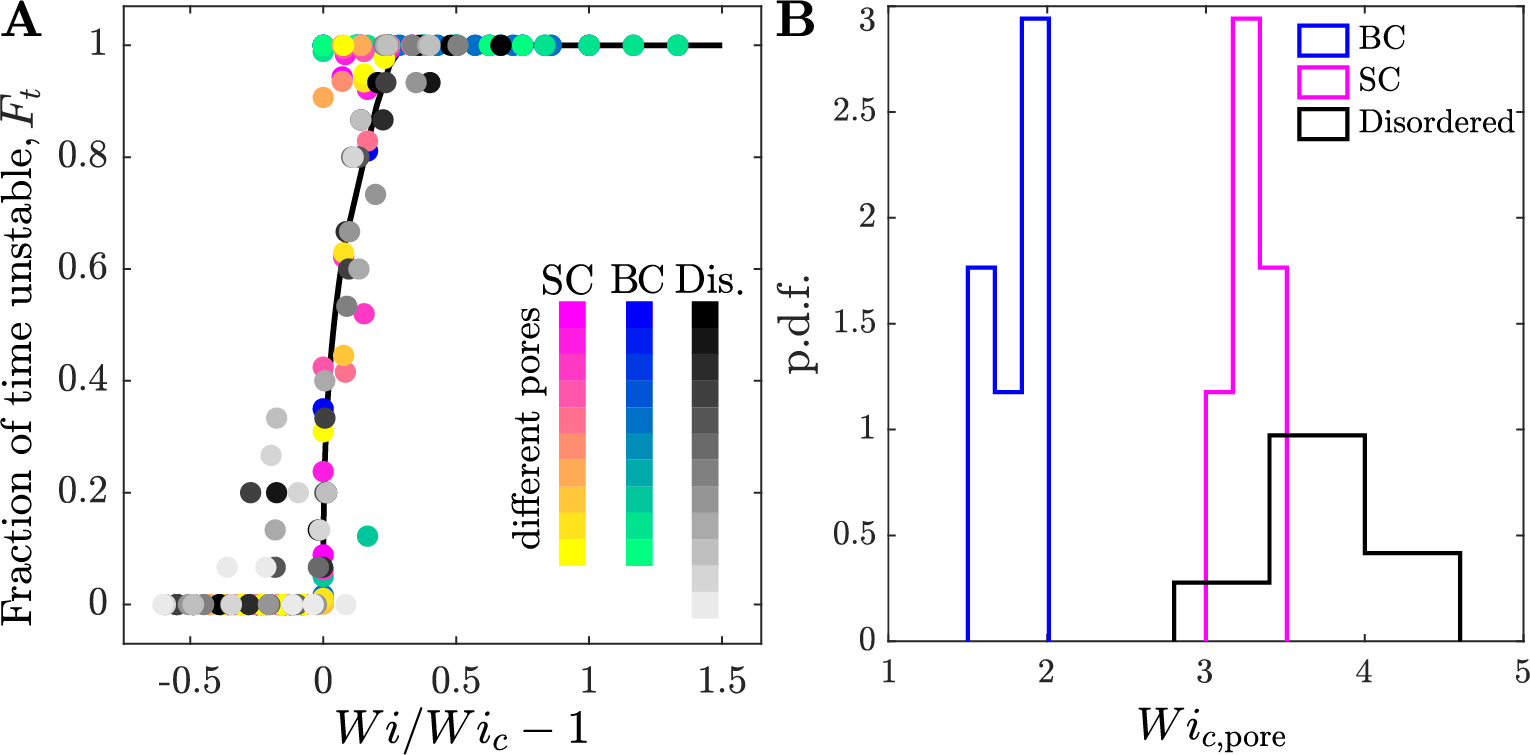}}
  \caption{Universal transition to elastic instability. (A) Fraction of time that a pore is unsteady ($F_t$) plotted against the macroscopic $Wi$ rescaled by $Wi_c$ for randomly selected pores across each porous medium, where the instability threshold criterion is $u'/|\mathbf{u}| > 0.2$. The transition to instability collapses for the ordered packings and disordered sphere packing. The solid black line corresponds to a square-root scaling and serves to guide the eye. (B) Distributions of the pore-specific critical Weissenberg number $Wi_{c,\mathrm{pore}}$ for randomly selected pores across each medium. The distributions are much narrower for the SC and BC geometries compared to the disordered packing, consistent with the hypothesis of porous individualism. Data corresponding to the disordered packing are reproduced from~\cite{browne_elastic_2021}.} 
  \label{transition}
\end{figure*}

Given that the framework of~\cite{mckinley_rheological_1996} provides a common way to describe the onset of the elastic instability across different geometries in our experiments, we next examine whether there may be any commonalities in how the flow transitions from steady to unsteady states. We characterize the extent of this transition by determining the fraction of time that pores are unsteady $F_t$, where values of $0<F_t<1$ correspond to transitory flow states with pores switching between steady and unsteady states over time. We image 10 randomly selected pores in each medium and apply an instability threshold of $u'/\langle |\mathbf{u}| \rangle_t > 0.2$ for at least 10\% of pixels in the frame to determine the instantaneous stability of each pore. This stability threshold was chosen to ensure that fluctuations calculated from the flow field are above the noise threshold resulting from particle image velocimetry. Figure~\ref{transition}A plots $F_t$ against the rescaled Weissenberg number for each pore, $Wi/Wi_c-1$. We also show data for the flow of an identical HPAM solution through a disordered random sphere packing with grain sizes of $D_p = 300-355 \: \upmu$m and porosity $\phi \approx 0.41$, reproduced from~\cite{browne_elastic_2021}. Remarkably, we find that the transition plots collapse not only for pores within the same geometry, but also across the three geometries overall, suggesting a common pathway to instability despite the different packing types. The solid black line corresponds to a square-root scaling and is shown to guide the eye.

Previously,~\cite{browne_elastic_2021} proposed the concept of ``porous individualism": pores of different geometries in a disordered porous medium can exhibit different flow states for the same imposed flow rate. This feature is reflected by a fairly broad distribution of threshold $Wi_{c,\mathrm{pore}}$ for individual pores sampled in a disordered packing, as shown in Figure~\ref{transition}B (reproduced from ~\cite{browne_elastic_2021}). By contrast, unlike the random disordered packing, all pores in the ordered packings studied here have nearly identical geometries. Indeed, we measure much narrower distributions of $Wi_{c,\mathrm{pore}}$ for pores sampled across each medium, as shown in Figure~\ref{transition}B. We hypothesize that the finite width of the distributions is due to polymer stress history effects, resulting from slight variations in the polymeric stress field from pore-to-pore. Further, we observe that the SC and disordered packing $Wi_{c,\mathrm{pore}}$ distributions overlap, while the BC distribution is located at overall lower $Wi_{c,\mathrm{pore}}$ values. It is still unclear what sets $Wi_{c,\mathrm{pore}}$ for each geometry, and investigating this point further will be an important direction for future work. However, we conjecture that this difference in onsets for the BC and SC geometries results primarily from a difference in the net stagnation point densities: the BC packing has a higher net overall density due to the presence of both inter-grain contact and upstream/downstream polar stagnation points.

\subsection{Flow topology reflects extensionally-dominated flow due to stagnation points}\label{flow_topology}
\begin{figure*}
\centerline{\includegraphics{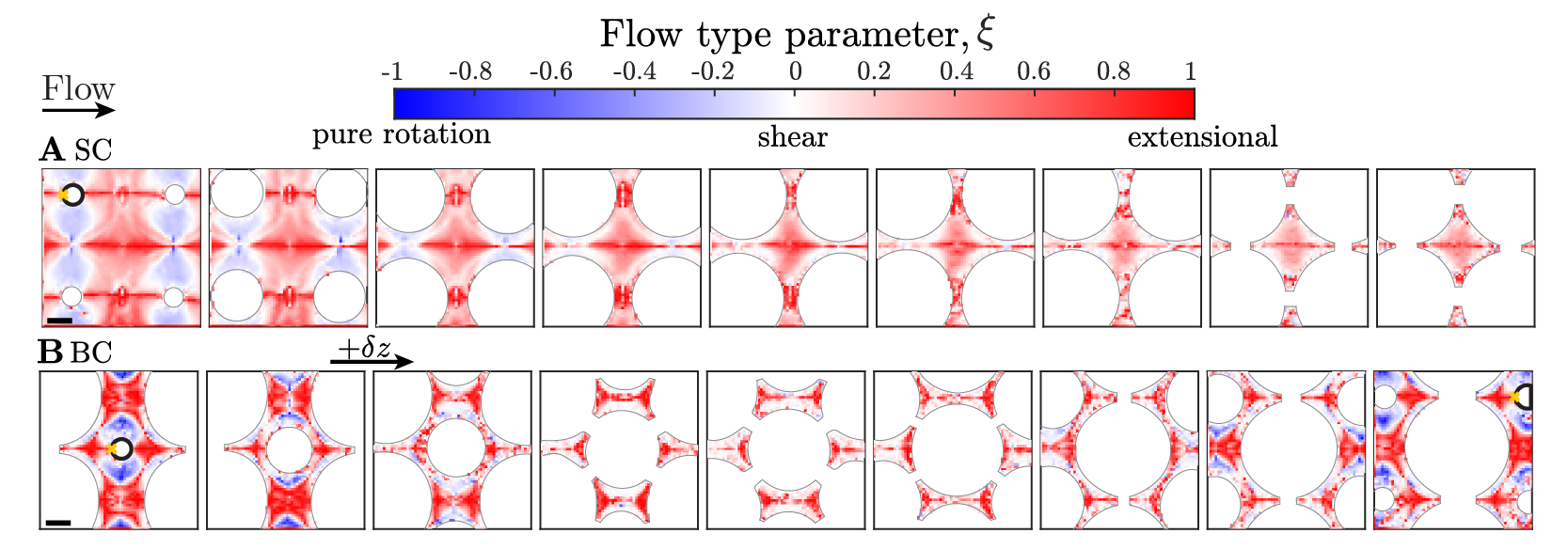}}
  \caption{Local flow type parameter ($\xi$) maps for the (A) SC and (B) BC packings at steady flow conditions ($Wi = 1.5$). Extensional flow is dominant near the inter-grain contacts and regions of the flow field with expansion and contraction of streamlines. Colormaps are presented for 9 $z$-planes spanning half of the characteristic unit cell volume, where sequential $z$-planes are spaced approximately $50 \: \upmu$m apart. Examples of contacts and contact-generated stagnation points are marked by black circles and yellow $\times$, respectively. The local $\xi$ fields are temporally averaged over 2 minutes of imaging to reduce noise. Scale bar is $200 \: \upmu$m.}
  \label{flowtype}
\end{figure*}

\begin{figure}
  \centerline{\includegraphics{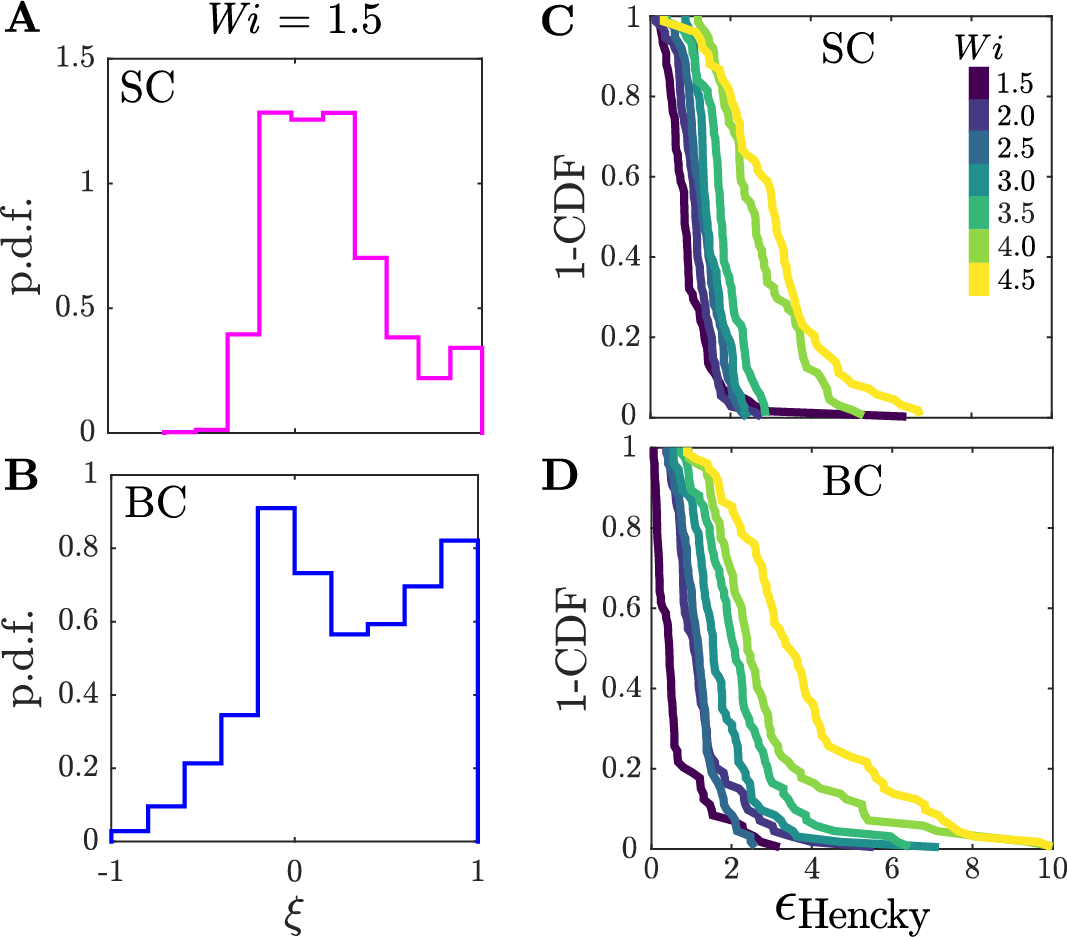}}
  \caption{Quantification of the extensional nature of the flow in the ordered packings. (A, B) Probability density functions (p.d.f.) of the flow type parameter, $\xi$, at $Wi = 1.5$ in the SC and BC packings. (C, D) Complementary cumulative distribution functions (1-CDF) of the Hencky strain, $\epsilon_{\mathrm{Hencky}}$, accumulated over 1 relaxation time across different Wi in the SC and BC packings. Appreciable values of the Hencky strain are accumulated in each geometry above the respective critical $Wi$, with the largest strains accumulating in the BC packing. Each distribution comprises 75 values.}
  \label{hencky}
\end{figure}

Our analysis thus far has assumed that stagnation points---which are dominated by locally extensile flow---are generated at inter-grain contacts, and these points generate the elastic instability at sufficiently large $Wi$. To directly confirm this assumption, we use the measured flow fields to determine the local flow type parameter $\xi=(|\mathbf{D}|-|\mathbf{\Omega}|)/(|\mathbf{D}|+|\mathbf{\Omega}|)$, where $|\mathbf{\Omega}|=\sqrt{2\mathbf{\Omega}:\mathbf{\Omega}}$ is the magnitude of the vorticity tensor~\citep{astarita_objective_1979,hamlington_direct_2008}. The flow type parameter takes on values from -1 to 1: $\xi=-1$ corresponds to solid-body rotation, $\xi=0$ corresponds to simple shear, and $\xi = +1$ corresponds to pure extension. Figure~\ref{flowtype} presents maps of $\xi$ for 9 $z$-planes spanning half of the characteristic unit cell volume at steady flow conditions ($Wi = 1.5$) in both packings, and Figure~\ref{hencky}A,B present full distributions of the measured values of $\xi$ that quantify the relative occurrences of flow type within the pore volume of each packing.

In the SC packing (Figure~\ref{flowtype}A), extensional flow dominates upstream and downstream of the inter-grain contacts, as expected. Extensional flow also exists in the middle of the bulk void space due to the contraction and expansion of the streamlines as the fluid navigates between successive pore bodies. These regions of local extension, arising from the stagnation points, are reflected by the skewed tail toward extensional flow shown in the measured distribution of $\xi$ (Figure~\ref{hencky}A). Similarly, in the BC packing (Figure~\ref{flowtype}B), extensional flow is found upstream and downstream of both the contacts and the upstream/downstream poles of the grains, while flow in the intermediate planes appears to be more shear-dominated due to the increased confinement. As a result, the flow topology distribution in the BC packing is more strongly skewed toward extensional flow ($\xi = 1$) with a peak also occurring near $\xi=0$ (Figure~\ref{hencky}B). Altogether, this in situ visualization directly confirms that inter-grain contacts generate extension-dominated stagnation points.

While the flow-type parameter can be used to identify regions of different flow topology, it does not provide information on the strength of each flow type. Instead, to quantify the strength of extensional effects, we calculate the Hencky strain accumulated by a polymer over the characteristic relaxation time using the local extension rate from the 2D velocity field, $\dot{\epsilon}=|\frac{\partial u}{\partial x}|+|\frac{\partial v}{\partial y}|$. The Hencky strain describes the net extensional deformation of a polymer chain due to an extensional velocity gradient over a certain time period; for sufficiently large Hencky strains, the polymers can undergo a coil-stretch transition~\citep{de_gennes_coilstretch_1974}, whereby the polymers become elongated and large extensional stresses develop~\citep{anna_interlaboratory_2001}. We select a starting location in the pore space and integrate forward in time along the Lagrangian trajectory to determine the accumulated Hencky strain, $\epsilon_{\mathrm{Hencky}}\approx \sum_{t=0}^{\lambda}\dot{\epsilon}\Delta t$. We repeat this for 5 random positions at 15 different time points to generate distributions of $\epsilon_{\mathrm{Hencky}}$ at different $Wi$. Figures~\ref{hencky}C and D present the resulting complementary cumulative distribution functions for the SC and BC packings. In both geometries, $\epsilon_{\mathrm{Hencky}}$ increases with $Wi$, reaching larger values in the latter compared to the former at matching $Wi$---reflecting the greater incidence of extension-dominated stagnation points in the BC packing. Moreover, for $Wi>Wi_c$ in both packings, $\epsilon_{\mathrm{Hencky}}$ predominantly exceeds 2---the value at which the extensional viscosity of dilute polymer solutions is known to drastically increase as a result of polymer stretching~\citep{anna_interlaboratory_2001,mckinley_filament-stretching_2002}. By contrast, for the flow of the same polymer solution in a disordered random sphere packing at similar values of $Wi$, the magnitudes of the accumulated Hencky strains are smaller than 1~\citep{browne_elastic_2021}. Therefore, we expect flow in the ordered packing packings to be more extensionally-dominant compared to more shear-dominated flow in a random sphere packing. As we show next, this difference in the flow type and resulting difference in polymer extension has marked implications for the macroscopic flow resistance through a porous medium.

\subsection{Flow thickening is generated by the elastic instability, primarily due to extensional effects}
Previous studies of HPAM solution flow resistance in porous rock cores~\citep{seright_new_2010,clarke_mechanism_2015,mitchell_viscoelastic_2016} and disordered 3D sphere packings~\citep{browne_elastic_2021} reported anomalous flow thickening---marked by a steep increase of the apparent viscosity $\eta_{\mathrm{app}}$ above the bulk shear viscosity $\eta_I$---for $Wi$ increasing above a threshold value.~\cite{browne_elastic_2021} demonstrated that this behavior reflects the onset of an elastic instability, and established that the increase in measured macroscopic flow resistance arises from the added viscous dissipation from  fluctuations in the unsteady flow field. However, their study was for a disordered medium, in which the flow is more shear-dominated, whereas our measurements here reveal that flow in \emph{ordered} 3D packings is more extensionally-dominant due to stagnation points generated at inter-grain contacts. What are the implications of this difference in flow type for the macroscopic flow resistance?

\begin{figure}
  \centerline{\includegraphics{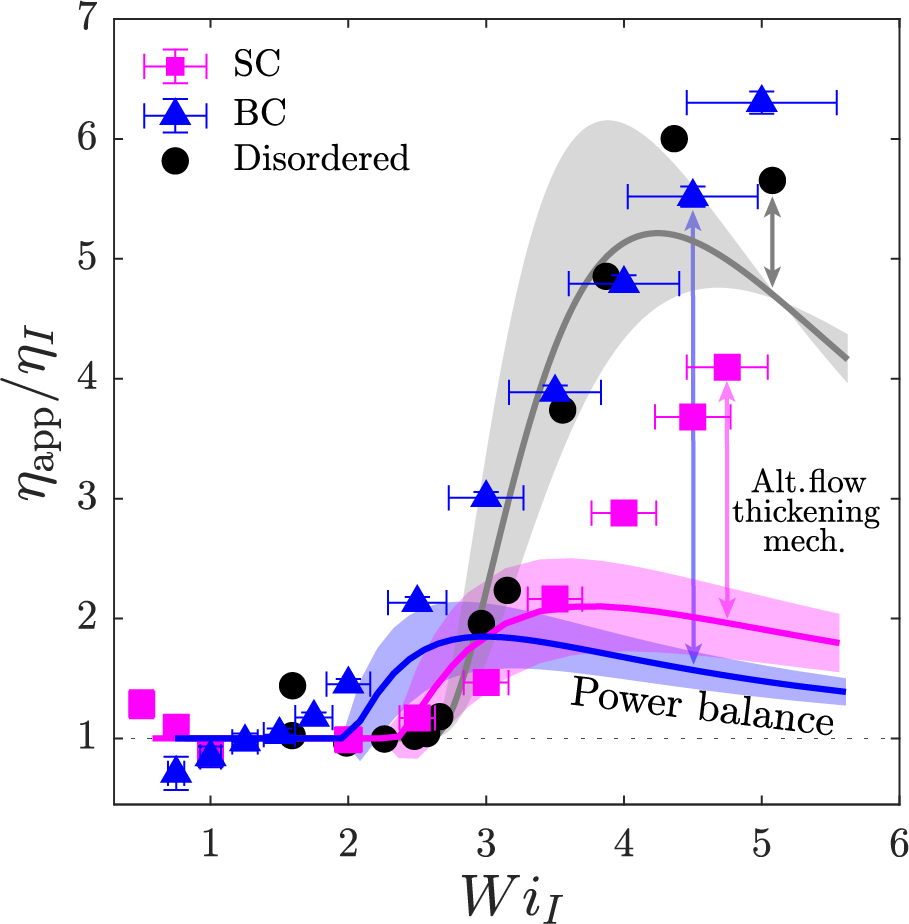}}
  \caption{Macroscopic flow resistance increases above a threshold $Wi$ for the same polymer solution in different 3D sphere packing geometries. The apparent viscosity $\eta_{\mathrm{app}}$ determined using Darcy's Law is normalized by the viscosity measured from bulk shear rheology $\eta_I$. The dotted line at $\eta_{\mathrm{app}}/\eta_I = 1$ is the Darcy flow resistance predicted by the bulk shear rheology. Solid curves correspond to the power balance model incorporating an added viscous dissipation generated by unsteady flow fluctuations. Shaded bands reflect the uncertainty in the power balance model. Data for the disordered packing are reproduced from~\cite{browne_elastic_2021}. The power balance captures the onset of flow thickening in the ordered packings but increasingly underpredicts the flow resistance with increasing $Wi$. Discrepancies between the power balance model and the measured flow resistance likely arises from extension-induced polymer stresses generated by contact-generated stagnation points in ordered packings, as indicated by the double-headed arrows. In some cases, error bars may be smaller than symbol size.}
  \label{powerbalance}
\end{figure}

To address this question, we measure the pressure drop across each medium as a function of $Wi$ and determine the corresponding apparent viscosity $\eta_{\mathrm{app}}$ using Darcy's law. We define a non-dimensional measure of flow resistance by normalizing $\eta_{\mathrm{app}}$ by the bulk shear viscosity $\eta_I$ evaluated at the corresponding interstitial shear rate. As shown in Figure~\ref{powerbalance}, both the SC and BC packings exhibit flow thickening above a threshold $Wi=Wi_c$, with $Wi_c\approx3$ and $Wi_c\approx2$ for SC and BC, respectively---coincident with the onset of the elastic instability (Figure~\ref{transition}B), as was found previously for disordered media as well. Our in situ flow visualization also provides a way to quantify the extent to which this flow thickening is due to the added fluid viscous dissipation from fluctuations generated by the instability, following the framework of~\cite{browne_elastic_2021}. Specifically, we directly compute the time-averaged rate of added viscous dissipation per unit volume due to pore-scale fluctuations from the flow field measured in single unit cells: $\langle \chi \rangle_{t,V}=\eta_I\langle\textbf{D}':\textbf{D}'\rangle_{t,V}$, where $\textbf{D}'$ is the rate-of-strain tensor associated with velocity fluctuations from the mean and we average over 9 $z$-planes spanning half of the characteristic unit cell volume. As detailed in~\cite{browne_elastic_2021}, balancing the rate of work done by the fluid pressure with the rate of viscous energy dissipation per unit volume yields an expression for how this added dissipation influences macroscopic flow resistance: 
\begin{equation}
    \eta_{\mathrm{app}}/\eta_I = 1 +k \langle \chi \rangle_{t,V}/(Q/A_c)^2 \eta_I.\label{thickening}
\end{equation}
This analysis assumes a power law fluid model that captures shear-thinning but neglects polymer elasticity. Therefore, it quantifies the additional power dissipated in the solvent due to flow fluctuations, but neglects the contribution of the polymeric stresses themselves, which we expect to be increasingly dominant in the ordered packings due to the greater prevalence of extensional flow.

This expectation is borne out by our data. Evaluating Eq.~\ref{thickening} using the measured flow fields produces the solid curves shown in Figure~\ref{powerbalance}. For both of the ordered packings, Eq.~\ref{thickening} captures the onset of flow thickening, but increasingly underpredicts the flow resistance at higher $Wi$---consistent with our expectation that extension-induced polymeric stresses play a more dominant role in these ordered sphere packings. Indeed, our local measurements of Hencky strain and flow type parameter highlight differences between extensionally-dominated flows in ordered packings and more shear-dominated flows in random, disordered packings, ultimately hinting at different mechanisms of increased flow resistance in ordered and disordered geometries. Moreover, the discrepancy in the power balance from the measured flow resistance is even larger for the BC packing, for which we found that extensional flow is even greater (Figures~\ref{flowtype} and~\ref{hencky}). Altogether, these data suggest that extensional viscosity arising from stagnation point-induced polymer extension has a larger contribution to the flow resistance in the SC and BC geometries, particularly with increasing $Wi$. Incorporating a spatially-varying polymer stress distribution into the model may improve the predictive power of Eq.~\ref{thickening}, and will be a useful direction for future work; however, in situ measurements of local polymer stress fields in porous media remain challenging experimentally.

\section{Conclusions} \label{sec:conclusions}
Our work has revealed that the spatially-extended contacts between grains in a 3D medium, which are typically idealized as discrete points and therefore overlooked, play a pivotal role in shaping the unsteady flow of a polymer solution through the pore space. They generate stagnation points that drive the onset of an elastic instability in both SC and BC sphere packings, contrasting the hypothesis that polar upstream and downstream stagnation points at sphere surfaces are the primary generator of elastic instabilities in flow past spheres. This finding suggests that not only is the effective stagnation point density within a particular geometry important, as suggested by~\cite{haward_stagnation_2021} for 2D pillar arrays, but that different classes of stagnation points may arise in flows through 3D porous media dependent on the porous medium geometry. Indeed, stagnation points arising at inter-grain contacts are inherently inaccessible in 2D model geometries, which many lab studies employ, yet are intrinsic to many natural porous media.

A simple rearrangement of the sphere packing type changes the pore space geometry confining the fluid, altering the distribution and density of stagnation points that contribute to the development of the elastic instability as well as the underlying flow topology. As a result, we expect that the features of the instability will be different for flow of the same polymer solution in the two different sphere packing geometries. Our in situ visualization confirmed this expectation: the instability generates fluctuating 3D eddies in the SC packing versus wobbling streamlines in the BC packing. Despite this marked difference in the pore-scale features of the unsteady flow between medium geometries, we found that there are common features of the transition pathway to instability across both the ordered and disordered porous media---suggesting that the route to elastic instability may be common across different pore structures.

~\cite{haward_viscosity_2003} and~\cite{odell_viscosity_2006} examined flow of dilute polymer solutions through model simple cubic and body-centered sphere packings, reporting an increase in macroscopic flow resistance that depends on the nature of the polymer, solution properties, and packing geometry. However, these and other previous characterizations of flow in 3D porous media were limited to quantifying bulk flow resistance, as access to in situ flow fields with high spatiotemporal resolution is typically hindered by the opacity of the medium. Our in situ flow visualization overcame this limitation and revealed the key role of inter-grain contact-generated stagnation points in controlling this flow thickening. In particular, we found that the steep increase in flow resistance coincides with the onset of local velocity fluctuations created by the elastic instability. However, unlike for flow in model pore constriction and expansion arrays~\citep{de_viscoelastic_2017-1,de_viscoelastic_2017,de_viscoelastic_2017-2} or disordered sphere packings~\citep{browne_elastic_2021,de_viscoelastic_2017-1}, added viscous dissipation arising from unsteady flow fluctuations does not fully capture the flow thickening behavior in the ordered packings. Rather, our data suggest a dominant role of polymeric stresses generated by the contact-associated stagnation points, which may manifest in complex ways~\citep{chauveteau_thickening_1984,evans_observations_1994,rothstein_axisymmetric_2001,haward_viscosity_2003,james_slow_2012,zamani_effect_2015,mokhtari_birefringent_2022}. Altogether, our results suggest that disordered porous medium geometries produce shear-dominated flows, consistent with a flow resistance largely governed by viscous dissipation in the solvent~\citep{browne_elastic_2021}, whereas ordered porous medium geometries generate more extensionally-dominated flow and thus elastic tensile stress-dominated flow resistance. Our current experimental methods lack the capability to quantify in situ polymer stress history; future work focusing on integrating rheological methods and simulations to incorporate stress history and holistically assess multiple mechanisms contributing to flow resistance will be important for fully understanding the mechanisms behind the anomalous flow thickening behavior of polymer solution flow in porous media.
    
In this work, we examined ordered porous media to help elucidate the relationship between 3D pore space geometry and viscoelastic flow behavior. Although random sphere packings provide a more realistic model of many porous media, it is more challenging to identify the relevant stagnation points in a random packing due to the irregular orientation of grains and pore-to-pore variations in mean flow direction. Investigating local flow behavior near inter-grain contacts in consolidated, random packings will be a useful direction for future work, as our results suggest that contact area-generated stagnation points have important implications for viscoelastic flow in random porous media. Further, a broad range of porous medium geometries exist in both natural settings and industrial applications beyond the binary classification of ordered versus disordered geometries. Our work focuses on granular-type packings containing consolidations modeled by neck-like contact areas, while other packing morphologies may include loose granular packings or media consisting of continuous solid matrices. We expect these different pore space geometries to fundamentally alter the flow behavior by controlling the flow topology and stagnation point densities. Indeed, for laminar Stokes flow—in the absence of instabilities altogether—variations in pore space geometry alone can control the onset of chaotic mixing of scalars~\citep{leborgne_lamellar_2015,lester_chaotic_2016,turuban_space-group_2018,turuban_chaotic_2019}. Examining polymer solution flow behavior in controlled model geometries of broader packing morphologies will thus be a useful extension of our work. 

By modeling features of naturally-occurring porous media in controlled microfluidic systems, our work provides an improved understanding of how pore space geometry influences the onset and features of viscoelastic flow instabilities in 3D porous media. Our results suggest the tantalizing possibility that elastic instabilities could be tailored via design of the pore space geometry for use in targeted applications, such as exploiting eddies to enhance solute mixing~\citep{browne_bistability_2020,chen_influence_2024} or leveraging wobbling streamlines to promote transverse dispersion of colloids~\citep{scholz_enhanced_2014}. Additionally, our results provide intuition for setting operating guidelines where viscoelastic porous media flows are relevant in which flow instabilities can either be prevented or utilized depending on application objectives.\\

\textbf{Funding.} We acknowledge support from the Princeton Center for Complex Materials (PCCM), a National Science Foundation (NSF) Materials Research Science and Engineering Center (MRSEC; DMR-2011750, as well as the Camille Dreyfus Teacher-Scholar Program of the Camille and Henry Dreyfus Foundation. S.J.H. and A.Q.S. gratefully acknowledge the support from the Okinawa Institute of Science and Technology Graduate University (OIST), which is funded by a subsidy from the Cabinet Office, Government of Japan. The authors also extend their thanks to the Engineering Section of the Core Facilities at OIST for their valuable assistance. This work was further supported by the Japan Society for the Promotion of Science (JSPS) through Kakenhi grants no. 24K07332 (S.J.H.) and no. 24K00810 (A.Q.S. and S.J.H.). \\

\textbf{Author contributions.} Conceptualization: C.A.B., S.J.H., S.S.D.; funding: S.J.H., A.Q.S., S.S.D.; experiments \& data analysis: E.Y.C.; contribution of experiment resources: S.J.H., A.Q.S., S.S.D.; interpretation of results: all authors; manuscript writing: E.Y.C., S.S.D; manuscript editing: all authors.\\

\textbf{Acknowledgements.} We thank K. Toda-Peters for fabricating the packing devices; D. W. Carlson, A. Clarke, and G. H. McKinley for insightful discussions; and the Princeton Imaging \& Analysis Center for use of the rheometer.

\newpage \appendix
\section*{Supplementary movie captions.}
\noindent SI videos are available from the corresponding author upon request.\\ 

\textbf{Movie 1.} Pathline imaging of flow behavior in the SC packing for increasing $Wi$ in the $z$-plane containing contacts between grains. Here, the contacts are the dark, circular regions in the field of view. Each panel corresponds to different applied flow rate conditions for the same imaging location over the course of one experiment. As $Wi$ increases above $Wi_c=3$, eddies form on the upstream faces of the contacts and grow in length with $Wi$ until reaching a highly fluctuating state. Pathlines are generated by averaging the mean pixel intensity over 30 successive frames. Videos play at 10$\times$ real time. Flow is from left to right. Scale bar is 200~$\upmu$m. 

\textbf{Movie 2.} Pathline imaging of flow behavior in the SC packing at $Wi = 1.5$ for 9 different $z$-planes spanning half of the unit cell. Each $z$-slice is spaced approximately 50~$\upmu$m apart. At $Wi = 1.5$, the flow is steady throughout the pore space. Pathlines are generated by averaging the mean pixel intensity over 30 successive frames. Videos play at 10$\times$ real time. Flow is from left to right. Scale bar is 200~$\upmu$m. Bright spots covering sphere surfaces are deposited tracer particles over the course of the flow experiment.

\textbf{Movie 3.} Pathline imaging of flow behavior in the SC packing at $Wi = 4.5$ for 9 different $z$-planes spanning half of the unit cell. Each $z$-slice is spaced approximately 50~$\upmu$m apart. At $Wi = 4.5$, the flow is unsteady due to the onset of an elastic instability, which manifests as highly fluctuating 3D eddies. Pathlines are generated by averaging the mean pixel intensity over 30 successive frames. Videos play at 10$\times$ real time. Flow is from left to right. Scale bar is 200~$\upmu$m. Bright spots covering sphere surfaces are deposited tracer particles over the course of the flow experiment.

\textbf{Movie 4.} Pathline imaging of flow behavior in the BC packing for increasing $Wi$ in the $z$-plane containing grain contacts. The larger dark regions correspond to the sphere bodies, while the smaller, circular dark region in the center is the contact between spheres. Each panel corresponds to different applied flow rate conditions for the same imaging location over the course of one experiment. The onset of the wobbling instability occurs above $Wi_c = 2$, where crossing pathlines initially develop upstream of the contact before fluctuations start to persist throughout the pore space. Pathlines are generated by averaging over 30 successive frames. Videos play at 10$\times$ real time. Flow is from left to right. Scale bar is 200~$\upmu$m. 

\textbf{Movie 5.} Pathline imaging of flow behavior in the BC packing at $Wi = 1.5$ for 9 different $z$-planes spanning half of the unit cell. Each $z$-slice is spaced approximately 50~$\upmu$m apart. At $Wi = 1.5$, the flow is steady throughout the pore space. Pathlines are generated by averaging the mean pixel intensity over 30 successive frames. Videos play at 10$\times$ real time. Flow is from left to right. Scale bar is 200~$\upmu$m. Bright spots covering sphere surfaces are deposited tracer particles over the course of the flow experiment.

\textbf{Movie 6.}
Pathline imaging of flow behavior in the BC packing at $Wi = 4.5$ for 9 different $z$-planes spanning half of the unit cell. Each $z$-slice is spaced approximately 50~$\upmu$m apart. At $Wi = 4.5$, the flow is unsteady due to the onset of an elastic instability, manifesting as strong transverse fluctuations which we term ``wobbling". Pathlines are generated by averaging the mean pixel intensity over 30 successive frames. Videos play at 10$\times$ real time. Flow is from left to right. Scale bar is 200~$\upmu$m. Bright spots covering sphere surfaces are deposited tracer particles over the course of the flow experiment.

\end{document}